\documentclass[aps,prd,twocolumn,showpacs,10pt,superscriptaddress,preprintnumbers,nofootinbib]{revtex4-1}
\usepackage{epsfig,amssymb,amsmath,psfrag,epstopdf,color}
\pdfoutput=1
\usepackage{graphicx}
\usepackage{subfig}
\usepackage[margin=7pt,justification=centerlast]{caption}

\usepackage{paralist}
\usepackage{hyperref}
\usepackage{float}
\usepackage{empheq}

\allowdisplaybreaks

\def\beq{\begin{equation}}
\def\eeq{\end{equation}}
\def\bsp#1\esp{\begin{split}#1\end{split}}
\newcommand{\be}{\begin{equation}}
\newcommand{\ee}{\end{equation}}
\newcommand{\bea}{\begin{eqnarray}}
\newcommand{\eea}{\end{eqnarray}}
\newcommand{\eqn}[1]{eq.~\eqref{#1}}

\def\Fig#1{Fig.~{\ref{#1}}}

\def\eqn#1{eq.~(\ref{#1})}

\def\App#1{Appendix~\ref{#1}}

\DeclareRobustCommand{\Sec}[1]{Sec.~\ref{#1}}

\def\cN{{\mathcal N}}
\def\cO{{\mathcal O}}
\def\cW{{\mathcal W}}

\def\to{\rightarrow}

\def\ksl{\not{\hbox{\kern-2.3pt $k$}}}

\def\e{\epsilon}

\def\Ord{{\cal O}}

\def\Re{{\rm Re}}

\def\bom#1{{\mbox{\boldmath $#1$}}}

\def\eqn#1{Eq.~(\ref{#1})}

\def\cN{\mathcal{N}}
\def\cO{\mathcal{O}}

\def\cQ{\mathcal{Q}}

\def\cW{\mathcal{W}}

\def\spa#1.#2{\left\langle#1\,#2\right\rangle}
\def\spb#1.#2{\left[#1\,#2\right]}
\def\lor#1.#2{\left(#1\,#2\right)}
\def\sand#1.#2.#3{%
\left\langle\smash{#1}{\vphantom1}^{-}\right|{#2}%
\left|\smash{#3}{\vphantom1}^{-}\right\rangle}

\def\lqcd{\Lambda_{\text{QCD}}}

\def\T{{\bom T}}

\newcommand{\nn}{\nonumber}

\def\supnu{{[\nu]}}

\def\tr#1{{(#1)}}

\newcommand{\two}{{(2)}}
\newcommand{\one}{{(1)}}

\begin{document}

\title{Rethinking Jets with Energy Correlators:\\Tracks, Resummation and Analytic Continuation}

\author{Hao~Chen}
\email{chenhao201224@zju.edu.cn}
\affiliation{Zhejiang Institute of Modern Physics, Department of
  Physics, Zhejiang University, Hangzhou, 310027, China\vspace{0.5ex}}
\author{Ian~Moult}
\email{imoult@slac.stanford.edu}
\affiliation{SLAC National Accelerator Laboratory, Stanford University, CA, 94309, USA\vspace{0.5ex}}
\author{XiaoYuan~Zhang}
\email{xyzhang0314@zju.edu.cn}
\affiliation{Zhejiang Institute of Modern Physics, Department of
  Physics, Zhejiang University, Hangzhou, 310027, China\vspace{0.5ex}}
\author{Hua~Xing~Zhu}
\email{zhuhx@zju.edu.cn}
\affiliation{Zhejiang Institute of Modern Physics, Department of
  Physics, Zhejiang University, Hangzhou, 310027, China\vspace{0.5ex}}

\begin{abstract}
We introduce an infinite set of jet substructure observables, derived as projections of $N$-point energy correlators, that are both convenient for experimental studies and maintain remarkable analytic properties derived from their representations in terms of a finite number of light ray operators. We show that these observables can be computed using tracking or charge information with a simple reweighting by integer moments of non-perturbative track or fragmentation functions. Our results for the projected $N$-point correlators are analytic functions of $N$, allowing us to derive resummed results to next-to-leading logarithmic accuracy for all $N$. We analytically continue our results to non-integer values of $N$ and define a corresponding analytic continuation of the observable, which we term a $\nu$-correlator, that can be measured on jets of hadrons at the LHC. This enables observables that probe the leading twist collinear dynamics of jets to be placed into a single analytic family, which we hope will lead to new insights into jet substructure. 
\end{abstract}

\maketitle

\tableofcontents
\phantom{x}
\newpage
\section{Introduction}
\label{sec:introduction}

The Large Hadron Collider~(LHC) provides a rich sample of high energy jets, opening up new opportunities to study the dynamics of QCD, and providing new avenues to search for physics beyond the Standard Model \cite{Asquith:2018igt,Larkoski:2017jix}. To perform first principles QCD calculations in the complicated environment of LHC collisions has required  significant theory progress, including the development of techniques to calculate groomed observables \cite{Larkoski:2014wba,Frye:2016aiz,Frye:2016okc}, and field theoretic formalisms for computing observables that incorporate the tracking \cite{Chang:2013iba,Chang:2013rca,Elder:2018mcr} or charge information \cite{Waalewijn:2012sv,Krohn:2012fg} often used to mitigate pile up and improve angular resolution. These advances have enabled the first comparisons of theoretical predictions with precision measurements for jet substructure observables \cite{TheATLAScollaboration:2013sia,Aad:2015cua,Sirunyan:2017tyr,Aaboud:2017qwh,Sirunyan:2018asm,Aad:2019vyi}.

Despite these successes, one of the drawbacks of observables that incorporate grooming algorithms or tracking information, is that this significantly complicates perturbative calculations, preventing the use of more modern techniques for loop and phase space integrals, and hindering the understanding of their underlying mathematical and field theoretic structure. This is particularly true for observables that use tracking information, which has prevented their use for precision measurements, despite their experimental advantages. To enable increasingly precise QCD measurements of jet substructure observables at the high luminosity LHC will require observables that are both amenable to higher order perturbative calculations, and that can be computed using tracking information. 

\begin{figure}
\includegraphics[width=0.785\linewidth]{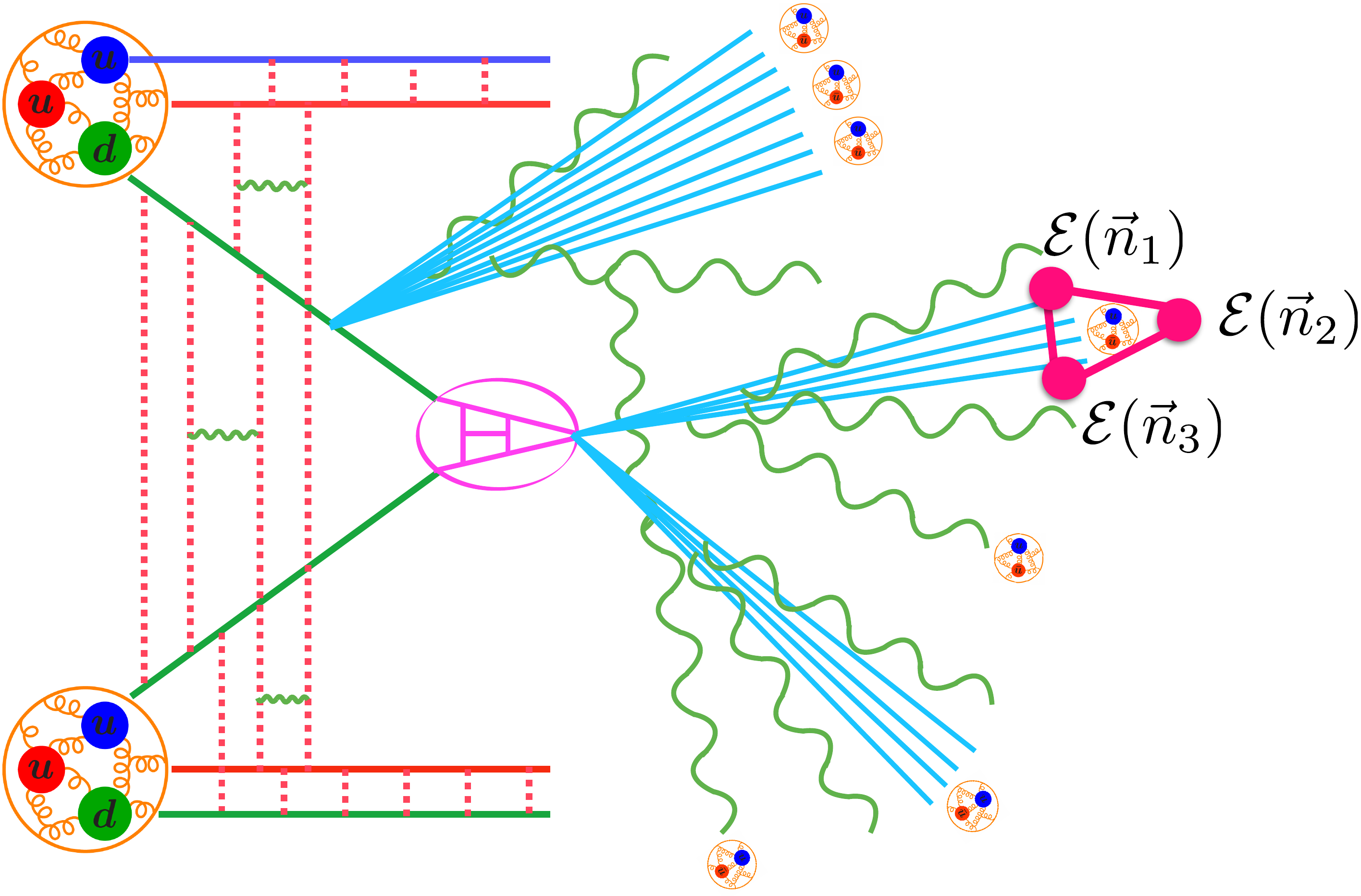}
\caption{ Energy flow operators, shown in red,  probe correlations between flows of energy arising from the collision of two protons at the LHC. In the small angle limit they factorize from the rest of the event and probe the collinear substructure of jets.
}
\label{fig:LHC}
\end{figure}

\begin{figure}
\includegraphics[width=0.685\linewidth]{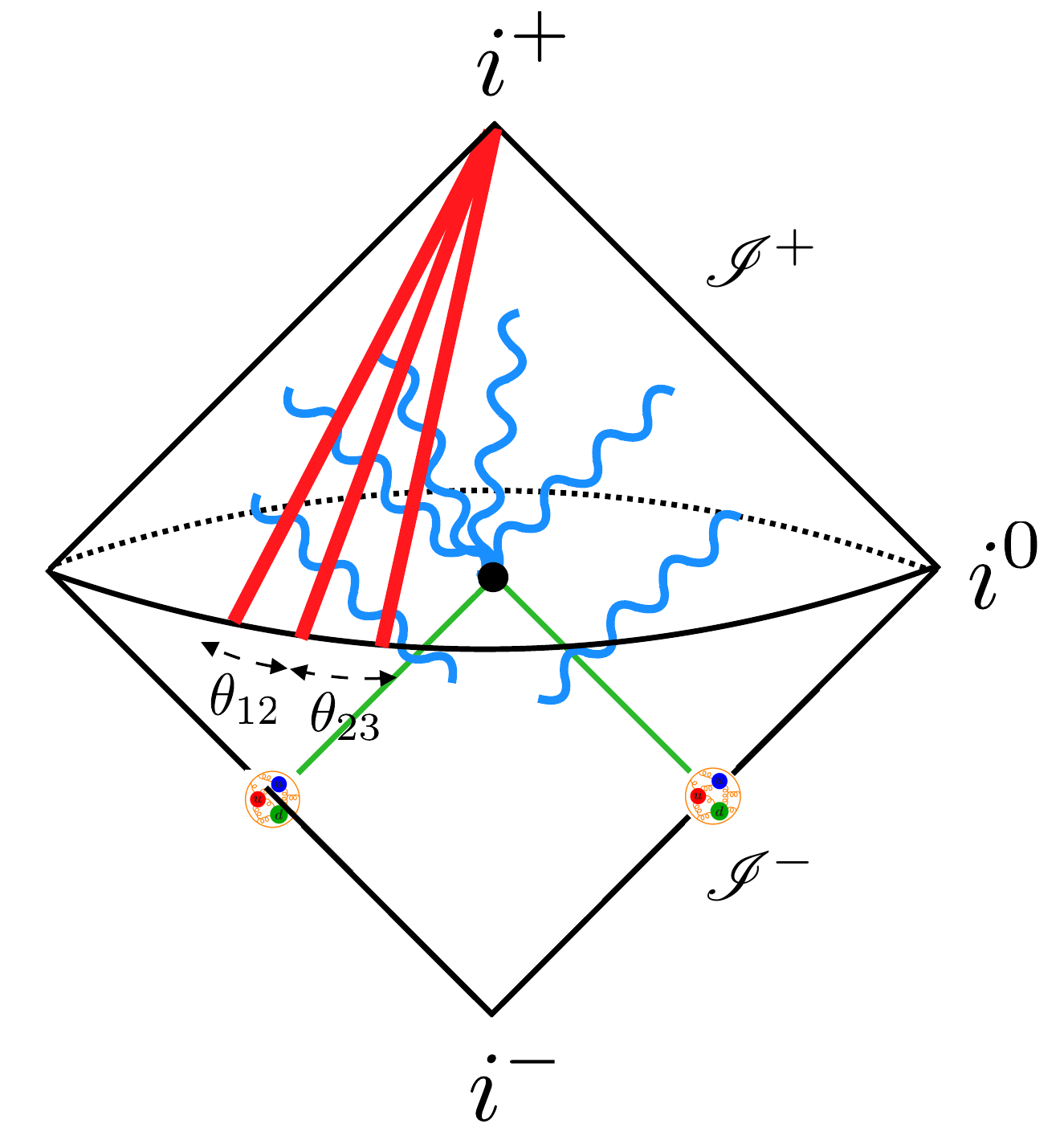}
\caption{Weighted cross sections can be formulated as matrix elements of a finite number of energy flow operators, leading to their simple theoretical properties.
}
\label{fig:penrose}
\end{figure}

While there has been significant effort towards the development of jet substructure observables at the LHC, it has primarily been from the perspective of developing tagging observables, rather than developing observables with the goal of simplifying their analytic structure. To understand what makes an observable simple from a theoretical point of view, one must begin by understanding what it means from a field theoretic perspective to measure the flow of energy (we will discuss later the case of charge) within a jet. The basic objects that measure energy flow are the energy flow operators \cite{Sveshnikov:1995vi,Tkachov:1995kk,Korchemsky:1999kt,Lee:2006nr,Hofman:2008ar,Belitsky:2013xxa,Belitsky:2013bja,Belitsky:2013ofa} defined as
\begin{align}\label{eq:ANEC_op}
\mathcal{E}(\vec n) =\int\limits_0^\infty dt \lim_{r\to \infty} r^2 n^i T_{0i}(t,r \vec n)\,,
\end{align}
where $\vec{n}$ is a unit three-vector that specifies the direction of the energy flow, and $T_{\mu\nu}$ is the energy-momentum tensor. 
The natural objects in the field theory are then correlation functions of these energy flow operators
\begin{align}\label{eq:correlator_intro}
\frac{1}{\sigma_{\rm tot}} \frac{d\sigma}{d \vec n_1 \cdots d\vec n_N}\stackrel{\rm F.T.}{=}\frac{\langle \cO \mathcal{E}(\vec n_1) \cdots \mathcal{E}(\vec n_N) \cO^\dagger \rangle }{\langle \cO  \cO^\dagger \rangle}\,,
\end{align}
which we will generically refer to as energy correlators.
In \eqn{eq:correlator_intro} the source operator $\cO$ in QCD can be, for example, the electro-magnetic current $\bar{\psi} \gamma^\mu \psi$, or Higgs operator $h/v G^{\mu\nu} G_{\mu \nu}$, and F.T. is a Fourier transformation to momentum space. Since we will not consider oriented observables in this paper, the Lorentz indices between $\cO^\dagger$ and $\cO$ can be contracted and will be ignored throughout. When all the energy flow operators in the correlator of \eqn{eq:correlator_intro} are placed in a collinear limit, these energy correlators are a jet substructure observable. This is illustrated for the particular case of a three particle correlator in \Fig{fig:LHC} from a particle physics perspective where the energy flow operators can be thought of as calorimeter cells, and in \Fig{fig:penrose} we show the spacetime structure of the energy flow operators in a Penrose diagram. However, as we will describe in detail in this paper, these energy correlators are quite distinct from the observables currently used for jet substructure at the LHC, largely due to the interests of the field during its developmental stages. For the particular case of two energy flow operators, the observable in \eqn{eq:correlator_intro} is referred to as the Energy-Energy correlator \cite{Basham:1978bw}, which has been used extensively as an $e^+e^-$ event shape (see e.g. \cite{Tulipant:2017ybb,Kardos:2018kqj} for recent work).

The energy correlator observables in \eqn{eq:correlator_intro} are in a sense the simplest observables in a field theory that measure the flow of energy. In particular, they inherit a number of simple theoretical properties from their direct representation as a matrix element:  they have manifest symmetry properties \cite{Belitsky:2013bja,Belitsky:2013xxa,Kologlu:2019mfz,Chen:2019bpb}, enjoy simple factorization properties in limits \cite{Moult:2018jzp,Gao:2019ojf,Dixon:2019uzg,Kologlu:2019mfz,Korchemsky:2019nzm,Chen:2019bpb}, have simple non-perturbative behavior even away from singular regions of phase space \cite{Korchemsky:1999kt},  can be analytically calculated to high perturbative orders \cite{Belitsky:2013ofa,Dixon:2018qgp,Luo:2019nig,Henn:2019gkr}, and can be directly studied using sophisticated techniques from conformal field theory (CFT) \cite{Kravchuk:2018htv,Kologlu:2019bco,Kologlu:2019mfz,Korchemsky:2019nzm}, including at strong coupling in $\cN=4$ super Yang-Mills (SYM) using the AdS/CFT correspondence \cite{Hofman:2008ar}. Furthermore, all infrared and collinear safe energy flow observables can be expressed in terms of these basic objects \cite{Sveshnikov:1995vi,Tkachov:1995kk} (for recent work see \cite{Komiske:2017aww,Komiske:2019asc}). While this connection is elegant, it is quite abstract, leading to a significant divide between the more formal theoretical study of simple energy correlator observables, and the ``real world" study of more experimental or phenomenological observables used at the LHC.

In this paper we attempt to bridge the theory-experiment divide by introducing observables that can be expressed in terms of correlation functions of a finite number of energy flow operators (as in \eqn{eq:correlator_intro}) and hence maintain simple theoretical properties enabling them to be computed to high perturbative orders, but that are simultaneously experimentally convenient.  We present the perspective that the simplest observables are precisely those that can be expressed in terms of correlation functions of a finite number of energy flow operators, and we ``give teeth" to this otherwise abstract perspective by concretely showing that it enables a number of new jet substructure calculations to higher perturbative orders, higher numbers of points, and incorporating tracking and charge information. We believe that this will have both an experimental impact, as well as make more transparent the connections between jet substructure and the more formal study of light ray operators. In this paper we will highlight a number of these advantages, leaving more phenomenological studies at higher perturbative orders, and with more detailed derivations, to future work.

In this paper we introduce the projected energy correlators, an \emph{infinite family} of experimentally convenient observables, each of which can be expressed in terms of a finite number of energy flow operators. These projected correlators behave similarly to common jet substructure observables such as the groomed jet mass, namely they are single logarithmic collinear (soft insensitive) observables designed to probe the collinear structure of jets. Furthermore, we show that this infinite family of observables in fact forms an analytic family, allowing us to derive results and perform resummation for arbitrary $N$-point projected correlators. 

One of the key benefits of the projected energy correlators that we will highlight in this paper is that they enable a simple incorporation of non-perturbative information relating to tracks or charges into perturbative calculations. The track function formalisms of \cite{Chang:2013iba,Chang:2013rca,Waalewijn:2012sv,Krohn:2012fg} have unfortunately not so far been widely applied for standard jet substructure observables, since such calculations are perturbatively complicated, and involve the full functional form of the non-perturbative track functions. In this paper, we show that the projected $N$-point correlators only require integer moments $\leq N$ that enter trivially as weights. Furthermore, the resummation of track correlators in the collinear limit only requires the renormalization of these integer moments, which satisfy linear renormalization group equations (as compared with the non-linear equation for the full track function), which enable them to be computed to higher perturbative orders. This will allow for high order perturbative calculations involving track information.

A further particularly interesting feature of our analysis is that our formulas for the $N$-point projected correlators are analytic functions of $N$ (for both the anomalous dimensions and the normalization constants), allowing us to consider their analytic continuation to non-integer values of $N$. These analytically continued observables have a scale evolution determined by the anomalous dimensions of non-integer twist-2 spin-$N$ operators.  We present a definition of these observables that is valid for measurements at the LHC. These observables correlate infinite combinations of particles within a jet (up to the fact that there are only a finite number of particles in real world applications). This illustrates a qualitatively new way of defining jet substructure observables through analytic continuation.  Analytic continuation also provides a means of defining families, in a mathematically precise sense, of observables that probe specific aspects of jets. In this language, one of the primary results of this paper is to place observables that probe the twist-2 dynamics of jets into a single analytic family.

An outline of this paper is as follows. In \Sec{sec:obs_v_weight} we discuss the difference between standard jet substructure observables and weighted cross sections, and emphasize that standard jet substructure observables necessarily involve matrix elements of an infinite number of energy flow operators. We then discuss the implications of this observation for incorporating track and charge information. In \Sec{sec:observables_a} we introduce projections of the energy correlators that are a function of a single scaling variable and are ideal for experimental studies. We also discuss ratios of these observables that are promising for precision measurements. In \Sec{sec:analytic} we analytically continue these observables to non-integer values of $N$, and define a new class of jet substructure observables which we term $\nu$-correlators. In \Sec{sec:N} we discuss the resummation of the $\nu$-correlators, and present numerical results for integer and non-integer values of $\nu$.  In \Sec{sec:N_track} we then generalize this to the case of correlators using tracking information. We conclude and discuss a number of future directions in \Sec{sec:conclusions}.

\section{Observables vs. Weighted Cross Sections}
\label{sec:obs_v_weight}

In this section we discuss the difference between standard ``observables" and ``weighted cross sections". In particular, we show that standard observables involve knowledge of an infinite number of energy correlators, and we will argue that weighted cross sections have a number of advantages, particularly when interfacing with non-perturbative data. A number of the properties of weighted cross sections and observables that are discussed in this section are known to experts in the field,\footnote{Unfortunately, we have found that different aspects are appreciated by non-overlapping groups of people.} however, we have chosen to discuss these issues in some detail since they are central to understanding the simplicity of the energy correlators. 

\begin{figure*}[t]
\centering
\subfloat[]{\label{fig:weighted}
\includegraphics[width=0.4\textwidth]{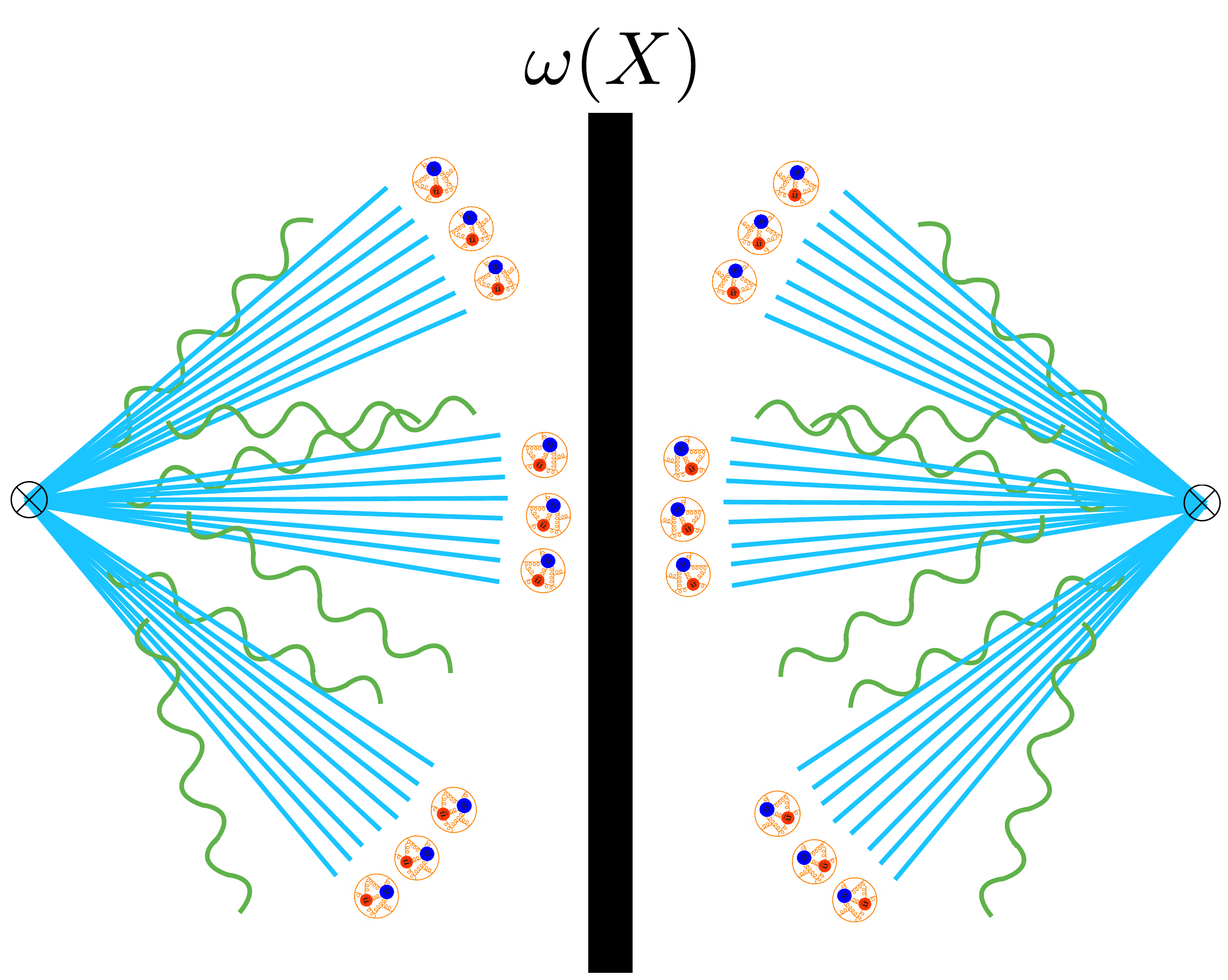}
}
\subfloat[]{\label{fig:delta}
\includegraphics[width=0.4\textwidth]{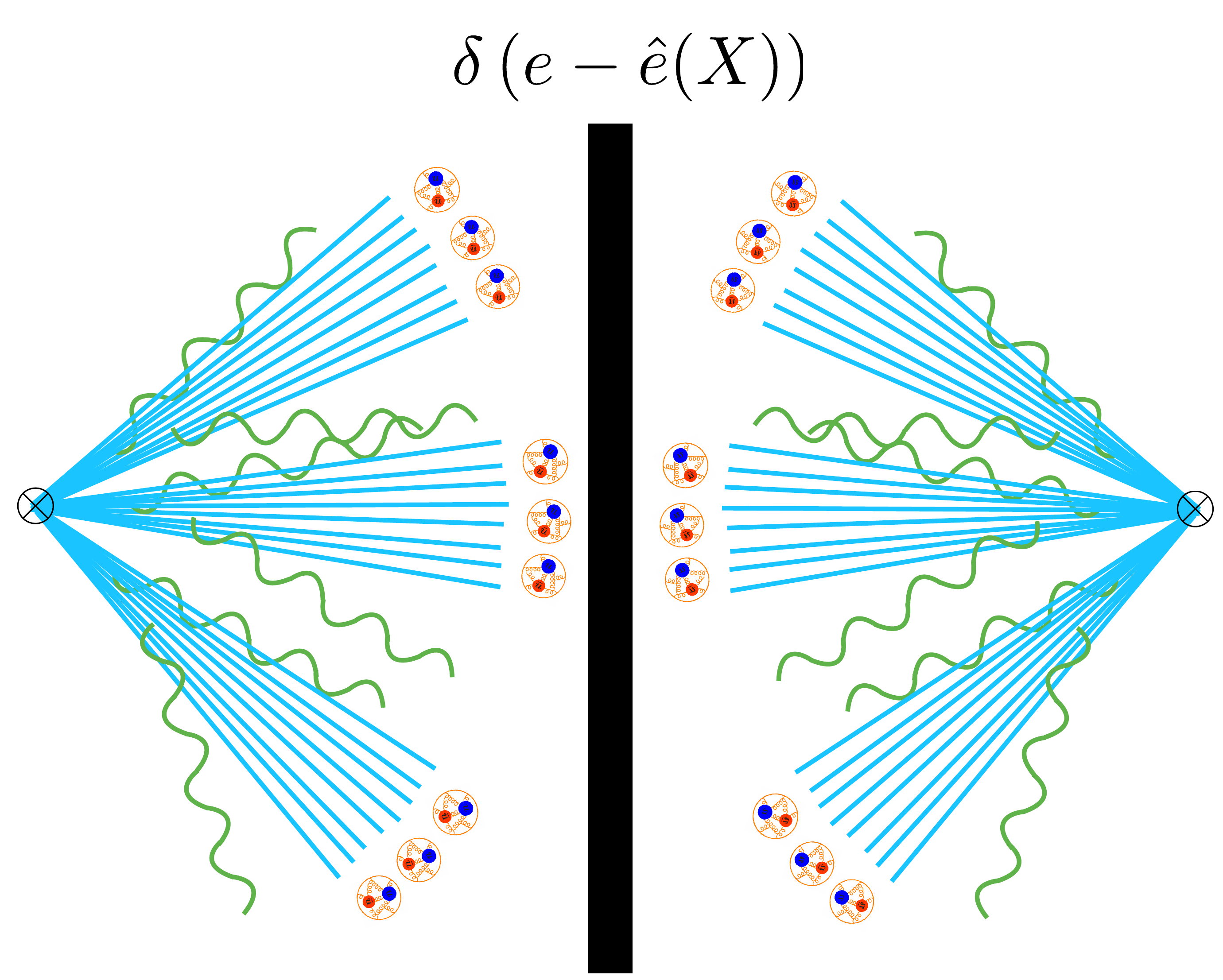}
}
\caption{An illustration of the difference between a weighted cross section, and a more standard jet observable. For a weighted cross section, shown in (a), a weighting function $\omega(X)$ is applied to the final state (here the cut is illustrated by the black bar). For a more standard jet observable, shown in (b), the final state is constrained by an operator $\hat e(X)$, and the cross section is calculated as a function of this constraint.}\label{fig:weighted_v_delta}
\end{figure*}

We begin by defining a weighted cross section as\footnote{This notation is borrowed from \cite{Belitsky:2013bja,Belitsky:2013ofa,Belitsky:2013xxa} where weighted cross sections in $\cN=4$ SYM were studied in detail, and whose perspective was influential to that presented here.}
\begin{align}
\sigma_\omega&= \int d^4x e^{iq\cdot x} \sum_{X}  \langle 0 | \cO (x) |X\rangle w(X) \langle X | \cO^\dagger(0) |0 \rangle \nn \\
&= \int d^4x e^{iq\cdot x}  \langle 0 | \cO (x)  \hat \omega \cO^\dagger(0) |0 \rangle\,.
\end{align}
Here $\hat \omega$ is a weight function that is a product of energy or charge flow operators that weights the asymptotic final state $X$, see \Fig{fig:weighted}, and $\omega(X)$ is the eigenvalue of $\hat \omega$ when acting on $X$. Restricting our attention for now to energy flow operators, we recall that the action of the energy flow operators in \eqn{eq:ANEC_op} on a state$| X \rangle$ is
\begin{align}
\mathcal{E}(\vec n)|X\rangle= \sum_i k_i^0 \delta^{(2)}(\Omega_{\vec{k}_i}-\Omega_{\vec n}) |X\rangle\,,
\end{align}
where $i$ runs over all particles in the state $| X \rangle$. We can therefore write an energy weight $\hat \omega$ as a product of energy flow operators
\begin{align}
\hat \omega=  \mathcal{E}(\vec n_1) \cdots \mathcal{E}(\vec n_N)\,.
\end{align}
This leads to an expression for an energy weighted cross section as a Wightman function
\begin{align}
\sigma_\omega&= \int d^4x e^{iq\cdot x}  \langle 0 | \cO (x)  \mathcal{E}(\vec n_1) \cdots \mathcal{E}(\vec n_N) \cO^\dagger(0) |0 \rangle\,.
\end{align}
Weighted observables are therefore directly expressible as matrix elements of energy flow operators. As mentioned in the introduction, the simple field theoretic definition of these objects has allowed significant recent progress in their understanding \cite{Chen:2019bpb,Moult:2018jzp,Dixon:2019uzg,Kologlu:2019mfz,Korchemsky:2019nzm,Chen:2019bpb,Belitsky:2013ofa,Dixon:2018qgp,Luo:2019nig,Henn:2019gkr}.  

Weighted cross sections defined in this manner are actually quite distinct from the observables that are most commonly used in jet substructure at the LHC. Instead of weighting the final state, it is more common to constrain it to have a particular value under the application of an operator $\hat e$,\footnote{We will often refer to such observables as ``$\delta$-function type observables".}
\begin{align}\label{eq:obs_def}
\frac{d\sigma}{de}= \int d^4x e^{iq\cdot x}  \langle 0 | \cO (x)  \delta(e-\hat e) \cO^\dagger (0) |0 \rangle\,,
\end{align}
as is shown schematically in \Fig{fig:delta}. Many familiar event shape observables (such as thrust \cite{Farhi:1977sg} or the angularities \cite{Berger:2003iw}) take this form, as do all jet substructure observables introduced for tagging purposes (such as $N$-subjettinesses \cite{Thaler:2010tr,Thaler:2011gf} or combinations of energy correlation functions \cite{Larkoski:2013eya,Larkoski:2014zma,Larkoski:2015kga,Larkoski:2014gra,Moult:2016cvt,Larkoski:2017iuy,Larkoski:2017cqq,Komiske:2017aww}). For these observables, the operator $\hat e$ can be expressed as an integral over energy flow operators with an angular weighting. Explicitly, for the case of thrust like event shapes which were studied in detail in \cite{Lee:2006fn,Bauer:2008dt,Mateu:2012nk}, one has
\begin{align}
\hat e |X\rangle =\frac{1}{Q}\int d\eta f_e(\eta) \mathcal{E}_T(\eta, \hat t)|X\rangle\,,
\end{align}
where 
\begin{align}
\mathcal{E}_T(\eta) = \frac{1}{\cosh(\eta)}\int d\phi \int\limits_0^\infty dt \lim_{r\to \infty} r^2 n^i T_{0i}(t,r \vec n)\,,
\end{align}
is the transverse energy flow operator, $Q = \sqrt{q^2}$,  $f_e(\eta)$ is an angular weighting function, and $\mathcal{E}(\eta, \hat t)$ is defined with respect to the thrust axis $\hat{t}$ and pseudo rapidity $\eta$. This construction extends in a straightforward manner to multi-particle correlations. For example, one can write the three particle correlations in small angle limit for jet substructure as
\begin{align}
e_3^{(\beta)}\propto\int d^2 \Omega_1 d^2 \Omega_2 d^2 \Omega_3 \mathcal{E}(\vec n_1) \mathcal{E}(\vec n_2)\mathcal{E}(\vec n_3) \theta_{12}^\beta \theta_{23}^\beta \theta_{31}^\beta \,,
\end{align}
where $\theta_{ij}$ denotes the angle between the vectors $\vec n_i$, $\vec n_j$.

While this definition of an observable may seem quite similar, the insertion of the $\delta$-function in \eqn{eq:obs_def} significantly complicates their structure relative to weighted cross sections. Unlike for weighted cross sections, which themselves can be directly expressed as matrix elements of energy flow operators, it is the moments of these observables
\begin{align}
\int e^n \frac{d\sigma}{de}= \int d^4x e^{iq\cdot x}  \langle 0 | \cO (x)  \hat e^n \cO^\dagger(0) |0 \rangle\,,
\end{align}
that are weighted cross sections. In particular, the moments of these weighted cross sections are directly related to the energy flow polynomials \cite{Komiske:2017aww}. The operator valued $\delta$-function in \eqn{eq:obs_def} is formally defined by its moments
\begin{align}
\delta(e-\hat e)=\delta(e)+\hat e \delta^{(1)}(e)+\cdots + \frac{\hat e^n}{n!} \delta^{(n)}(e)+\cdots\,,
\end{align}
and observables of this form therefore require the knowledge of correlators of an infinite number of energy flow operators.  In particular,  we conclude that  any observable that is defined by specifying its value on the final state involves an infinite number of energy correlators to define it~(again, up to the fact that there are only a finite number of particles in real world applications).

In this paper, we want to advocate that the use of weighted cross sections provides many advantages, particularly in the context of precision calculations. The fact that standard observables involve an infinite sum over all moments hints that they are sure to be a more complicated object, and are likely to obscure the simple symmetry properties of the underlying energy correlators. While in perturbation theory, this is perhaps acceptable,\footnote{Although we should emphasize that the perturbative simplicity of energy correlator observables has enabled a number of analytic calculations \cite{Belitsky:2013ofa,Dixon:2018qgp,Luo:2019nig,Henn:2019gkr,Chen:2019bpb} that were not possible for standard $\delta$-function observables, leading to valuable perturbative data for improving our understanding of event shapes \cite{Moult:2019vou}.} we will see that this complication is particularly transparent when considering non-perturbative effects such as the inclusion of track information. In particular, we will show that observables involving only a finite number of energy correlators is particularly convenient, will require only a finite number of moments of non-perturbative functions, instead of an infinite number. This allows for new calculations of track based observables, and is one of the key points that we wish to emphasize in this paper.

Although it is not the primary goal of this paper, it is also worth emphasizing that the nature of the physics being probed by the ``weighted observables" such as the energy correlators, as compared with $\delta$-function type observables is actually quite different. In particular, energy correlators are by definition probing energy correlations at a particular angular scale. This ensures that they probe the collinear core of a jet, and are insensitive to wide angle soft radiation. This is quite distinct from having a constraint $\delta (e-\hat e)$ and demanding $e \ll 1$ as is commonly done in jet substructure. Due to the energy weighting necessary in the observable for infrared and collinear safety, this condition is also satisfied by soft radiation, giving rise to soft sensitivity. There has been much interest in the jet substructure community in achieving observables that are insensitive to soft radiation, primarily focused on starting with observables that are soft sensitive and eliminating this sensitivity by grooming. However, the restriction to collinear physics can be automatically achieved by starting with weighted cross sections, and we believe that this perspective is beneficial from a theoretical perspective. 

Finally, we conclude this section with a comment on the adoption of ``observables" as opposed to weighted cross sections in the study of jet substructure at the LHC. The rejuvenation of the study of the dynamics of QCD jets was largely driven by the search for beyond the Standard Model physics, and in particular, the construction of jet observables that tag jets with particular energy flows. Unlike standard ``observables", weighted cross sections do not take a single value on a given jet. For example, for the two point energy correlator, each pair of particles within the jet gives an entry into the distribution, as opposed to a single entry from the jet itself. Therefore weighted cross section type observables are not by themselves obviously useful for tagging.\footnote{Although, as mentioned above, their moments are directly related to the energy flow polynomials which are a basis of tagging observables \cite{Komiske:2017aww}. It would also be interesting to understand how to use weighted cross sections in the search for new physics. For an early example of an observable that is closely related to the energy correlators being used for new physics searches, see \cite{Jankowiak:2011qa}.} As jet substructure has transitioned to the precision study of QCD properties, the same observables originally used for tagging have continued to be used. However, as we will argue in this paper, in the context of precision measurements, we should completely reconsider the classes of observables that are used in the study of jet substructure, and we will show that energy correlators offer a number of significant advantages.

\subsection{Incorporating Tracks}
\label{sec:obs_v_weight_tracks}

\begin{figure}[t!]
\includegraphics[width=0.885\linewidth]{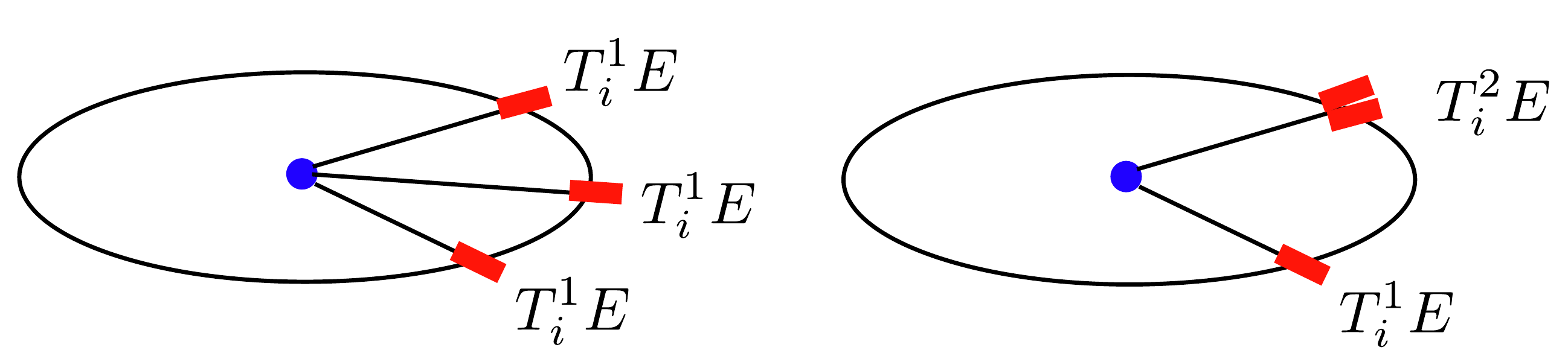}
\caption{Energy correlators using tracks. a) When the detectors are widely separated, only the first moment of the track functions appear, and simply rescales the weighting function. b) Higher moments of the track functions appear in contact terms when the two detectors are placed at the same angle. These contact terms are necessary for describing collinear limits.
}
\label{fig:tracks}
\end{figure}

One of the key advantages of weighted cross sections that we highlight in this section is that they interface in a simple manner with tracking information. This should be intuitive: instead of weighting by the total energy flowing in a particular direction, one must simply change to weighting by the energy flowing in tracks in that direction. This modification only requires the knowledge of a single (measurable) non-perturbative number, the average energy converted into tracks, see \Fig{fig:tracks}. The goal of this section is to make this precise using the language of track functions. The results of this section hold for generic angles between the energy correlators, and are not restricted to the collinear limit. The collinear limit will be considered in more detail in \Sec{sec:N_track}, and here we will find additional simplifications that arise when considering resummation with tracks.

In \cite{Chang:2013iba,Chang:2013rca} an elegant field theoretic formalism for the treatment of tracks was developed\footnote{See also \cite{Elder:2017bkd} for a generalization of the track function and jet charge formalism to fractal observables.} that allows for the separation of perturbative and non-perturbative physics through the introduction of a track function $T_i(x)$, with $i$ denoting the parton label, $i=q\,,g$. The precise field theoretic definition of the track function is not required here. It describes the distribution in energy fraction of a parton $i$ that hadronizes into tracks (charged particles) with four momentum $\bar p_i^\mu=x p_i^\mu +\cO(\lqcd)$. Here $0\leq x \leq 1$ and the track function satisfies the sum rule
\begin{align}
\int\limits_0^1 dx~T_i(x,\mu)=1\,.
\end{align}
The track function is a non-perturbative object, but has a calculable scale ($\mu$) dependence, similar to a fragmentation function. 
We will define the following shorthand notation for the moments of the track function
\begin{align}
T_i^\tr{n}= \int\limits_0^1 dx~ x^n \, T_i(x,\mu)\,.
\end{align}
At the level of detail that we work to in this section, one can imagine that to convert a perturbative calculation to a calculation on tracks, one must simply tack a track function onto each parton \cite{Chang:2013iba,Chang:2013rca}.  However, we will see that this process is much simpler for weighted cross sections as compared to $\delta$-function type observables.

We first consider the case of an observable defined with a $\delta$-function
\begin{align}
\frac{d\sigma}{de}= \sum\limits_N \int d\Pi_N \frac{d\sigma_N}{d\Pi_N} \delta \left[   e-\hat e(\{ p_i^\mu \}) \right]\,,
\end{align}
where we use $d\sigma_N$ to denote the $N$-body differential cross section, and $d \Pi_N$ the $N$-body Lorentz invariant phase space measure. 
The observable defined on tracks is then given by
\begin{align}\label{eq:track_obs}
\frac{d\sigma}{d \bar e}= \sum\limits_N \int d\Pi_N \frac{d \bar\sigma_N}{d\Pi_N}   \int \prod\limits_{i=1}^N dx_i\, T_i (x_i)\delta \left[   e-\hat e(\{x_i  p_i^\mu \}) \right]\,.
\end{align}
Here we have followed the notation of \cite{Chang:2013rca} where the bar over the observable indicates the observable measured on tracks. In \eqn{eq:track_obs}, $d \bar\sigma_N/d\Pi_N$ denotes a matching coefficient. In general, the analytic calculation of observables on tracks is complicated because the measurement constraint now involves the variables $x_i$. This is not only a technical complication, but as we will see shortly, it will also imply that the observable depends on the complete functional form of the non-perturbative track function.

On the other hand, for an energy correlator it is trivial to incorporate tracking information, since this just rescales the weight function. This is shown schematically in \Fig{fig:tracks}.
For a particular partonic configuration (and for well separated correlators), the conversion to tracks is achieved by making the following replacement for the weights
\begin{align}\label{eq:replace_charge}
E_i \to \int dx_i \, x_i T_i(x_i) E_i =T^\tr{1}_i E_i\,.
\end{align}
In other words, in going to a calculation in tracks, the first moment of the track function appears as a multiplicative constant for the weight, either $T_q^\tr{1}$ or $T_g^\tr{1}$~($T_q^\tr{n} = T_{\bar q}^\tr{n}$ due to the charge conjugation invariance of QCD). This means that at any loop order one can trivially convert partonic calculations for the energy correlators to calculations on tracks. The moments of the track functions can then be directly measured in experiment.

As an example to illustrate the difference in complexity between these two situations, we consider the LO calculation for both the thrust observable, which is a standard observable of the form of \eqn{eq:obs_def}, and the two-point energy correlator (EEC). The LO calculation for thrust was presented in \cite{Chang:2013iba}, 
\begin{align}
\frac{d\sigma}{d\bar \tau}&=\int\limits_0^1 dy_1 dy_2 \frac{d\bar \sigma(\mu)}{dy_1 dy_2} \int\limits_0^1 dx_1 dx_2 dx_3 T_q(x_1)T_q(x_2) T_g(x_3) \nn \\
&\delta\left[ \bar \tau -\bar \tau(y_1, y_2, x_1, x_2, x_3)  \right]\,,
\end{align}
with
\begin{align}
\frac{d\bar \sigma(\mu)}{dy_1 dy_2}=\sigma_0 \frac{\alpha_s(\mu) C_F}{2\pi} \frac{\theta(y_1+y_2-1)(y_1^2+y_2^2) }{(1-y_1)(1-y_2)}\,,
\end{align}
where $y_1 = 2 E_q/Q$, $y_2 = 2 E_{\bar q}/Q$ are the normalized parton energy, and the measurement function for track thrust is
\begin{align}
  \label{eq:track_thrust_meas}
  \bar \tau = &\ \theta[x_1 x_3 (1-y_2) - x_1 x_2 (1-y_3)] 
\nn\\
&\
\cdot \theta[x_2 x_3 (1-y_1) - x_1 x_2 (1-y_3) ] x_1 x_2 ( 1 - y_3) 
\nn\\
&\
+
\theta[x_2 x_3 (1-y_1) - x_1 x_3 (1-y_2)] 
\nn\\
&\
\cdot \theta[x_1 x_2 (1-y_3) - x_1 x_3 (1-y_2) ] x_1 x_3 ( 1 - y_2) 
\nn\\
&\
+
\theta[x_1 x_3 (1-y_2) - x_2 x_3 (1-y_1)] 
\nn\\
&\
\cdot \theta[x_1 x_2 (1-y_3) - x_2 x_3 (1-y_1) ] x_2 x_3 ( 1 - y_1)  \,,
\end{align}
where $y_3 = 2 - y_1 - y_2$.
Already at LO, one can see that this calculation is non-trivial, and the result involves the complete functional dependence on the non-perturbative track functions. This also makes it complicated to interface with numerical calculations performed using subtraction schemes.

On the other hand, for the EEC, the calculation at LO is trivial, since it simply involves weighting the contribution from the correlation of two quarks by $(T_q^\tr{1})^2$ and the contribution from the correlation of a quark and a gluon by $T_q^\tr{1} T_g^\tr{1}$. For an $e^+e^-$ source, we find
\begin{align}
\text{EEC}(z)=\sigma_0 \frac{\alpha_s}{2\pi} C_F \left( (T^\tr{1}_q)^2 I_1(z) +2T^\tr{1}_q T^\tr{1}_g I_2(z) \right)\,,
\end{align}
where
\begin{align}
I_1&=\left( \frac{1}{6z^2}+\frac{1}{z^3}-\frac{4}{z^4}  \right) \frac{1}{1-z}+\left( \frac{3}{z^4}-\frac{4}{z^5}  \right)\frac{\ln(1-z)}{1-z}\,, \nn \\
I_2&=\left( \frac{53}{12z^2}-\frac{41}{4z^3}+\frac{13}{2z^4}\right) \frac{1}{1-z} \nn \\
&+\left(  \frac{13}{2z^5}-\frac{7}{z^4}+\frac{2}{z^3}  \right) \ln(1-z)\,,
\end{align}
and $z = (1 - \cos\theta)/2$ is the angle between the two correlated partons. 
This calculation involves no additional complexities as compared to the standard fixed order calculation, and only requires knowledge of the first moments of the track functions, which are numbers (not functions). Calculations beyond LO are possible using the ingredients of the ordinary EEC calculation~\cite{Dixon:2018qgp,Luo:2019nig}.

To deal with collinear limits, as illustrated in \Fig{fig:tracks}, one must also consider the placement of multiple correlators on the same parton. For an $N$-point correlator, one must consider up to $N$ correlators placed on a single parton. If $N$ correlators are placed on the same parton, one gets the $n$-th moments of the track functions
\begin{align}
E^n_i \to \int dx_i~ x_i^n\, T_i(x_i) E^n_i =T^\tr{n}_i E^n_i\,.
\end{align}
These higher moments will be required when we consider the resummation of the track energy correlators in the collinear limit.

To illustrate the presence of these higher moments, we can consider the gluon jet function for the EEC in the collinear limit. This was computed without tracking information through two loops in \cite{Dixon:2019uzg}. For the differential jet function,\footnote{Later, we will also need the integrated jet function, which is simply the cumulant of the differential jet function.} the scale-independent piece was found to be
\begin{align}
j_g(z)&=\delta(z)+ \frac{\alpha_s}{4\pi} \left( \frac{14}{5}C_A +\frac{1}{5}n_f   \right)  \left[  \frac{1}{z} \right]_+   \nn \\
&+\delta(z)   \frac{\alpha_s}{4\pi}  \left( -\frac{898}{75}C_A -\frac{14}{25}n_f   \right)\,.
\end{align}
On tracks the result is simply
\begin{align}
&j_g(z)=\delta(z) T_g^\tr{2} \\
&+ \frac{\alpha_s}{4\pi} \left( \frac{14}{5}C_A (T_g^\tr{1})^2  +\frac{1}{5}n_f (T_q^\tr{1})^2   \right) \left[  \frac{1}{z} \right]_+  \nn \\
&+\delta(z)   \frac{\alpha_s}{4\pi}  \left( -\frac{898}{75}C_A (T_g^\tr{1})^2  -\frac{14}{25}n_f (T_q^\tr{1})^2  \right) + \Ord(\alpha_s^2) \,. \nn
\end{align}
This result is intuitive, in particular the second moments $T_g^\tr{2}$ and $T_q^\tr{2}$ appear as the coefficients of the leading order $\delta(z)$ contact terms, while the terms with a non-trivial $z$ dependence are weighted by $(T_q^\tr{1})^2$ or $(T_g^\tr{1})^2$, arising from the detectors being placed on distinct particles, following the replacement rule in \eqn{eq:replace_charge}.

We can again compare this to a track based calculation for an observable defined via a $\delta$-function constraint. Even for a ``simple" observable such as thrust, the one-loop jet function contains complicated dependence on the track functions. We do not reproduce the full result here, which can be found in Eq. (48) of \cite{Chang:2013iba}, but illustrate just a couple terms in the result
\begin{widetext}
\begin{align}
\bar J(\bar s, x, \mu) \supset
\frac{\alpha_s C_F}{2\pi} \int \limits_0^1 dx_2 \int \limits_0^1 \frac{dz}{z}  & T_q \left( \frac{x-(1-z)x_2}{z} \right) T_g(x_2) \left[  \frac{1}{\mu^2} \mathcal{L}_0 \left(  \frac{\bar s}{\mu^2} \right) (1+z^2) \mathcal{L}_0(1-z)  \right.  \\
&\left. +\delta(\bar s)\left((1+z^2)\mathcal{L}_1(1-z) +\ln \left(  \frac{xz^2}{[x-(1-z)x_2]x_2}  \right)(1+z^2)\mathcal{L}_0(1-z)+1-z \right)   \right]\,. \nn
\end{align}
\end{widetext}
Here $\mathcal{L}_0$ and $\mathcal{L}_1$ are standard plus functions.
This result involves the complete functional dependence of the track functions, which are non-perturbative objects. It also suggests that calculations at two, or three loops would be quite complicated. From our perspective, this complicated dependence is easily understood as arising from the fact that the thrust observable requires knowledge of an infinite number of correlators.

One final comment is in order. It is standard to consider cross sections where the weight is a conserved quantity. Here of course, the track number, or the ``energy in tracks", is not conserved, nor is it related to a combination of standard Noether charges. Nevertheless, we can formally define a charged energy flow operator as
\begin{align}
\mathcal{T}_c(\vec n) =\int\limits_0^\infty dt \lim_{r\to \infty} r^2 n^i   T^c_{0i}(t,r \vec n)\,,
\end{align}
with the action on a state is
\begin{align}
\mathcal{T}_c(\vec n)|X\rangle= \sum\limits_{i\in X_c} k_i^0 \delta^{(2)}(\Omega_{\vec{k}_i}-\Omega_{\vec n}) |X\rangle\,.
\end{align}
Here $T^c$ is the stress tensor but only involving the charged fields, and similarly $X_c$ denotes the charged particles in the state $X$. This gives energy correlators measured on tracks a clean field theoretic definition. 

It would be interesting to compute the full angle EEC on tracks at higher perturbative orders. This is straightforward analytically at NLO following the calculations of \cite{Dixon:2018qgp,Luo:2019nig}, since going from a standard perturbative calculation to a calculation on tracks simply requires tagging partonic configuratiosn with appropriate track functions. It could also be performed numerically at NNLO using standard subtraction schemes.\footnote{The EEC observable has been calculated numerically at NNLO using the ColorfulNNLO subtraction scheme \cite{DelDuca:2016ily,DelDuca:2016csb}.} As with the standard EEC, there exists a sum rule relating the integrated EEC cross section to the total energy in tracks, which can be computed perturbatively \cite{Chang:2013rca}. Such a calculation could be interesting for attempting to resolve potential discrepancies for $\alpha_s$ extractions from event shapes, both by providing an additional handle, and since the experimental data for track observables is significantly more precise.

\subsection{Incorporating Charges}
\label{sec:obs_v_weight_charge}

Although we will not discuss it in much detail in this paper, it should also be immediately clear that we can extend the above discussion of tracks to the calculation of charge correlators.  Here we will consider the measurement of the object $E\cdot Q$. i.e. we define the operator
\begin{align}
\mathcal{Q}_1(\vec n) =\int\limits_0^\infty dt \lim_{r\to \infty} r^2 n^i  J_0(t,r \vec n)T_{0i}(t,r \vec n)\,,
\end{align}
where the notation follows that in \cite{Waalewijn:2012sv}, namely that the subscript $1$ indicates the energy weighting. The energy weighting is convenient experimentally, and will also keep the renormalization group evolution identical to that of the energy correlators. Charge correlators without the energy weighting are also interesting, and have been studied in detail in $\mathcal{N}=4$ SYM \cite{Belitsky:2013ofa,Belitsky:2013bja,Belitsky:2013xxa}, and at leading order in QCD \cite{Chicherin:2020azt}, but we will not study them here.

The one point correlators $\langle \mathcal{Q}_1(\vec n_1)\rangle$ which measure the average charge of the jet, as well as the two point contact term $\langle \mathcal{Q}_1(\vec n_1) \mathcal{Q}_1(\vec n_1) \rangle$, which measures the width of the jet charge distribution on a jet, have both been studied in  \cite{Waalewijn:2012sv,Krohn:2012fg}, and have been measured \cite{TheATLAScollaboration:2013sia,Aad:2015cua,Sirunyan:2017tyr}. In fact, the entire jet charge distribution has been measured, so in principle all the moments are known. Like the track functions these objects are non-perturbative, but their renormalization group evolution can be computed perturbatively \cite{Waalewijn:2012sv}.

One can now  study multi-point correlators of these objects, $\langle \mathcal{Q}_1(\vec n_1) \mathcal{Q}_1(\vec n_2)\cdots \mathcal{Q}_1(\vec n_N)\rangle$, or correlators with some standard energy operators stuck in. Here, just as with the track functions, the angular dependence can be computed perturbatively, and the only non-perturbative inputs that are required are integer moments of the appropriate non-perturbative functions. For example for the two-point correlator, one needs the following moments of the fragmentation functions
\begin{align}
\tilde D_i^Q&=\sum \limits_h Q_h \int \limits_0^1 dz ~z D_{hi}(z,\mu)\,,  \\
\tilde D_i^{Q^2}&=\sum \limits_h Q^2_h \int \limits_0^1 dz ~z^2 D_{hi}(z,\mu)\,.
\end{align}
In QCD one has the relations
\begin{align}
\tilde D_g^Q&=\sum \limits_h Q_h \int \limits_0^1 dz ~z D_{hg}(z,\mu)=0\,,
\end{align}
and
\begin{align}
\tilde D_q^{Q^n}&=(-1)^n\tilde D_{\bar q}^{Q^n}\,,
\end{align}
These were extracted from various parton shower programs in \cite{Waalewijn:2012sv}. For a generic $N$-point correlator, one also needs
\begin{align}
\tilde D_i^{Q^n}&=\sum \limits_h Q^n_h \int \limits_0^1 dz ~z^n D_{hi}(z,\mu)\,.
\end{align}
With these, one can then immediately algorithmically weight partons in perturbative calculations to obtain correlators of $\cQ_1$. We will not discuss these objects further in this paper, but we think they would be interesting to measure, and calculate to higher orders. They probe interesting correlations well beyond what have been studied previously.

\section{Experimental Observables}
\label{sec:observables_a}

Having illustrated the simple properties of the energy correlators, one may be under the impression that they are a fairly constrained set of observables. For example, the two point energy correlator by itself is a single observable (unlike the angularities \cite{Berger:2003iw} one cannot add an angular weighting to their definition), and higher point correlators become increasingly complicated functions of multiple variables that are not easily amenable to experimental analyses. To overcome this, the goal of this section is to introduce an infinite family of experimentally convenient observables that depend on a finite number of energy correlators.

\subsection{Projected Energy Correlators}
\label{sec:observables_proj}

The simplest class of observables are distributions of a single scaling variable. We would therefore like to generalize the two point correlator to obtain scaling variables that probe complementary aspects of the collinear structure of jets.

The simplest extension of the two point correlator is to consider higher point correlators, but integrate out all the information about the shape keeping the longest side fixed. This effectively determines the size of the $N$ points being measured. We should point out that this is not the only way to integrate out information. A different possibility is to find the diameter of the minimal enclosing circle of the $N$ points being measured, and use this diameter as the scaling variable. We will refer to these observables as ``projected $N$-point correlators". In this paper we will consider the longest side definition only, and will later show how to generalize this definition to non-integer values of $N$, which we will refer to as $\nu$-correlators. Studying the dependence on the longest side gives access to the scaling behavior of higher point correlators. We will begin by defining these observables in $e^+e^-$ collisions, where they are defined for generic angles.
We will then consider their restriction to the collinear limit, where they can be defined on jets at the LHC. Throughout this section we will provide definitions in both a continuum (or CFT) language, as well as in a particle language applicable for experimental measurements at the LHC.

\begin{figure}
\includegraphics[width=0.645\linewidth]{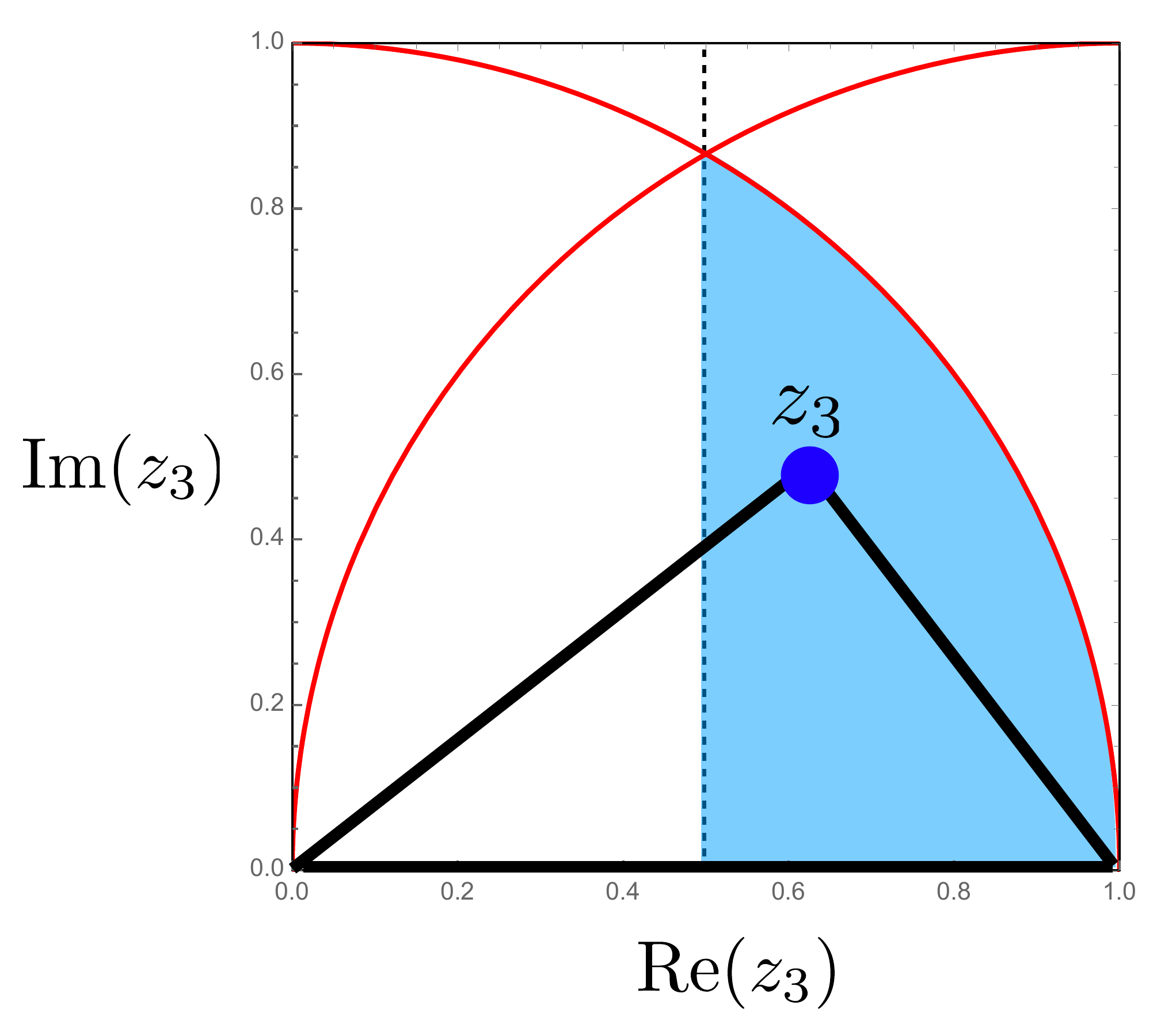}
\caption{The geometry of the three point correlator: the longest side is placed along the real axis, and the third point of the triangle, $z_3$ lies inside the shaded blue region. The region to the left of the dashed line is related by parity. To define the projected correlator, $z_3$ is integrated over the shaded blue region.
}
\label{fig:integration_region}
\end{figure}

We define the projected $N$-point correlator as
\begin{align}
  \label{eq:projection}
\hspace{-0.5cm}  &\frac{d\sigma^{[N]}}{d x_L} \ = \int\! d\Omega_{\vec{n}_1} 
\! \int\! d\Omega_{\vec{n}_2} 
\delta (x_L - \frac{1 - \vec{n}_1 \cdot \vec{n}_2}{2} ) 
\prod_{k=3}^N \int \!  d\Omega_{\vec{n}_k}
\nn\\
&\ 
\Theta(\{\vec{n} \}) \int\! d^4 x\, \frac{e^{i q \cdot x}}{Q^N} \langle 0 | \cO^\dagger(x)   
{\cal E}(\vec{n}_1) {\cal E}(\vec{n}_2)  \ldots 
{\cal E}( \vec{n}_N)
\cO (0) | 0 \rangle \,,
\end{align}
where
\begin{equation}
\label{eq:areaint}
d \Omega_{\vec n} = \frac{1}{4 \pi} \sin\theta d \theta d \phi\,, 
\end{equation}
is the area element on the celestial sphere. The integration region for $d\Omega_{{\vec n}_k}$ is specified by
\begin{align}
  \label{eq:omega12k}
  \Theta(\{ \vec{n} \})  = \prod_{
\substack{1\leq i<j\leq N
\\
i+j > 3 
}
} \theta(|\vec{n}_1 - \vec{n}_2|  - |\vec{n}_i - \vec{n}_j|) \,,
\end{align}
namely, we fix the angular distance between the first two energy flow operators to be $x_L = (1 - \cos\theta_{12})/2$, and integrate over the remaining operators with the constraint that their mutual angular distance, as well as their angular distance with respect to the first two operators, to be smaller than $x_L$.
Taking the concrete case of the projected three point correlator, this definition involves integrating the three point correlator (whose analytic form was computed at LO in \cite{Chen:2019bpb}) over the configuration space of three points shown in \Fig{fig:integration_region}. 
The integration over ${\cal E}(\vec{n}_k)$, $k > 2$ will lead to contact terms when two or more energy flow operator are placed at the same point in the celestial sphere. Such terms are straightforward to deal with in the momentum space factorization language in $D = 4 - 2 \e$ dimension~\cite{Dixon:2019uzg}. In particular, the integration over the area is non-singular, as is guaranteed by the average null energy condition.\footnote{The integration of the energy flow operator over a small area element gives the energy $\omega$ deposited in that area. The ANEC states that $\omega$ is always semi-positive, and the finiteness of total energy implies that $\omega$ is finite.} The finiteness of the integration ensures the infrared and collinear (IRC) safety of the observable. 

By definition the projected $N$-point correlators have support for  $x_L \in [0, 1]$, and obey the sum rule
\begin{align}
  \label{eq:sumrule_N}
  \int_0^1 \! dx_L \,  \frac{d\sigma^{[N]}}{d x_L}  = \sigma_{\rm tot} \,,
\end{align}
which follows from 
\begin{align}
\sum_{1 \leq i_1 , \ldots , i_N \leq n} \frac{\prod_{a=1}^N E_{i_a}}{Q^N} = \frac{(\sum_{i=1}^n E_i )^N}{Q^N} = 1 \,.
\end{align}
This is an extension of the sum rule for the two point case \cite{Dixon:2019uzg,Korchemsky:2019nzm,Kologlu:2019mfz}.

We can also define the projected $N$-point correlators on a discrete set of particles. This is more convenient for experimental measurements and perturbative calculations in momentum space. Suppose we have a scattering with center-of-mass energy $Q$ into $n$ particles, $\{p_{1}\,, p_{2}\,, \ldots \,, p_{n} \}$. The projected $N$-point correlator can then be calculated as
\begin{align}
  \label{eq:projected_mom}
  \frac{d\sigma^{[N]}}{d x_L} &\ = \sum_n \sum_{1 \leq i_1,\ldots, i_N \leq n } 
\int\! d\sigma_{e^+e^- \to X_n}
\frac{\prod_{a=1}^N E_{i_a}}{Q^N}
\nn\\
&\ \cdot \delta(x_L - \max \{x_{i_1 i_2}, x_{i_1 i_3}, \ldots , x_{i_{N-1} i_N } \}) \,,
\end{align}
where $X_n$ denotes a $n$-particle final state and $x_{ij} = (1 - \vec{n}_i \cdot \vec{n}_j)/2 = (1 -  \cos\theta_{ij})/2$ is the two-particle angular distance. The summation over $n$ is needed to ensure IRC safety, and in the second sum, we allow the $i_a = i_b$ term, which corresponds to the contact term mentioned before.  The $\delta$-function picks out the largest angle separation in the $N (N-1)/2$ angles. Eq.~\eqref{eq:projected_mom} applies non-perturbatively. In perturbation theory, $X_n$ consists of asymptotic quarks and gluons, while non-perturbatively it consists of discrete hadrons. For $N=2$ this reduces to the usual definition of the EEC. 

In a simulation or experiment, \eqn{eq:projected_mom} can be implemented as follows. For a scattering event consisting of $n$ final-state particles, the weight in the bin $[x_L - \Delta, x_L + \Delta]$ is given by~
\begin{align}
  \label{eq:int_weight}
  W_\Delta(x_L)&\  = \frac{1}{2 \Delta} \sum_{1 \leq i_1,\ldots, i_N \leq n } \frac{\prod_{a=1}^N E_{i_a}}{Q^N}
\\
&\ \cdot \theta(\max \{x_{i_1 i_2}, x_{i_1 i_3}, \ldots , x_{i_{N-1} i_N } \} - (x_L - \Delta) )
\nn\\
&\ \cdot \theta(x_L + \Delta - \max \{x_{i_1 i_2}, x_{i_1 i_3}, \ldots , x_{i_{N-1} i_N } \} ) \,. \nn
\end{align}
The full histogram is obtained by filling all the bins, summing over all events, and dividing by the total number of events.

\begin{figure}[ht!]
  \centering
  \includegraphics[width=0.42\textwidth]{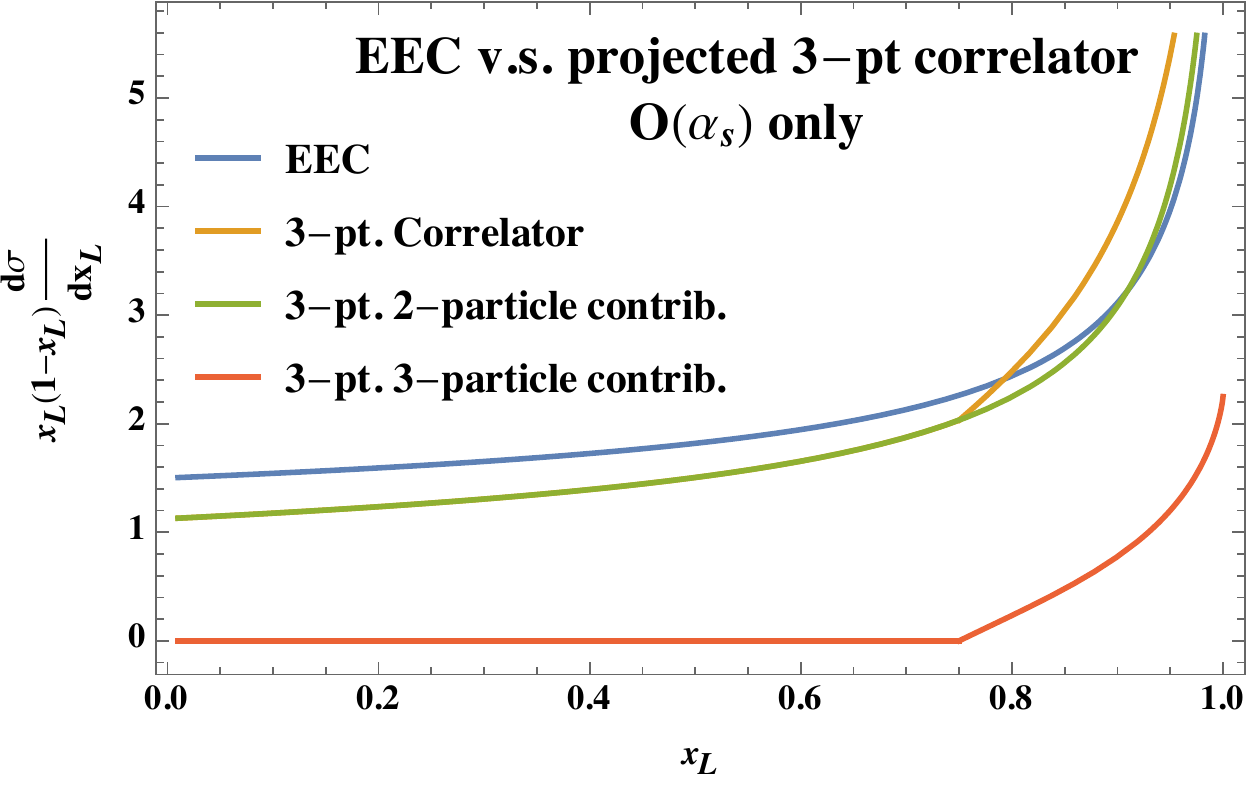}
  \caption{Comparision between the standard EEC and the projected $3$-point correlator at $\cO(\alpha_s)$. Plotted here is the coefficient of $\sigma_0 C_F \alpha_s/(4 \pi)$. We have also multiplied the distribution by $x_L (1-x_L)$ to suppress the collinear~($x_L \to 0$) and back-to-back~($x_L \to 1$) singularities. The remaining singular behavior at $x_L \to 1$ is due to Sudakov double logarithms.}
  \label{fig:eec_vs_3pt}
\end{figure}

At lowest order in perturbation theory, it is straightforward to calculate the projected $N$-point correlator analytically for generic angles. As an example, for $N=3$ and $0 < x_L < 1$, we have for $e^+e^-$ annihilation
\begin{align}
  \label{eq:sig_N_eq3}
  \hspace{-0.4cm}\frac{d\sigma^{[3]}}{d x_L} &= \sigma_0 \frac{\alpha_s}{4 \pi} C_F \Bigg[
-\frac{3 \left(20 x_L^3-75 x_L^2+87 x_L-30\right) \ln (1-x_L)}{2 (1-x_L) x_L^6}
\nn\\
&\ -\frac{3 \left(8
   x_L^3-83 x_L^2+144 x_L-60\right)}{4 (1-x_L) x_L^5}
+ \theta(x_L - \frac{3}{4})
\nn\\
&\
\cdot
\Bigg(
-\frac{3 \left(-256 x_L^4+1264 x_L^3-2088 x_L^2+1407 x_L-333\right)}{8 (1-x_L) x_L^6}
\nn\\
&\
-\frac{3
   \left(8 x_L^4-56 x_L^3+123 x_L^2-105 x_L+30\right) \ln (1-x_L)}{(1-x_L) x_L^6}
\nn\\
&\
+ \frac{6 \left(8 x_L^3-48 x_L^2+75 x_L-30\right) \ln (2)}{x_L^6}
\Bigg) \Bigg] + \cO(\alpha_s^2) \,,
\end{align}
where the first two lines are due to contributions where two (and only two) indices in $i_1\,, i_2\,, i_3$ are identical in Eq.~\eqref{eq:projected_mom}~($2$-particle contribution), while the term proportional to the step function $\theta(z - 3/4)$ is due to the contribution where $i_1 < i_2 < i_3$~($3$-particle contribution). The point with $x_L = 3/4$ comes from the Mercedes-Benz configuration, where the pair-wise angle is $2 \pi/3$. This is the fully symmetric configuration for a three particle final state. This number will decrease order by order in perturbation theory, and for a perfectly spherical symmetric radiation pattern will reduce to $0$. In Fig.~\ref{fig:eec_vs_3pt}, we plot the projected $3$-point correlator at $\cO(\alpha_s)$ (weighted by $x_L (1-x_L)$ to suppress the contact term), along with the separate $2$- and $3$-particle contributions. As a comparison we also show the result for the standard EEC.

The projected correlators are particularly convenient in the collinear limit where the non-analytic behavior (e.g. the $\theta$-function in \eqn{eq:sig_N_eq3}) that is present for generic angles is power suppressed. For jets at the LHC, one can simply define the identical observable, but restricted to the constituents of a jet identified using some jet algorithm.

\begin{figure*}[ht!]
\centering
\subfloat[]{\label{fig:pi}
\includegraphics[width=0.3\textwidth]{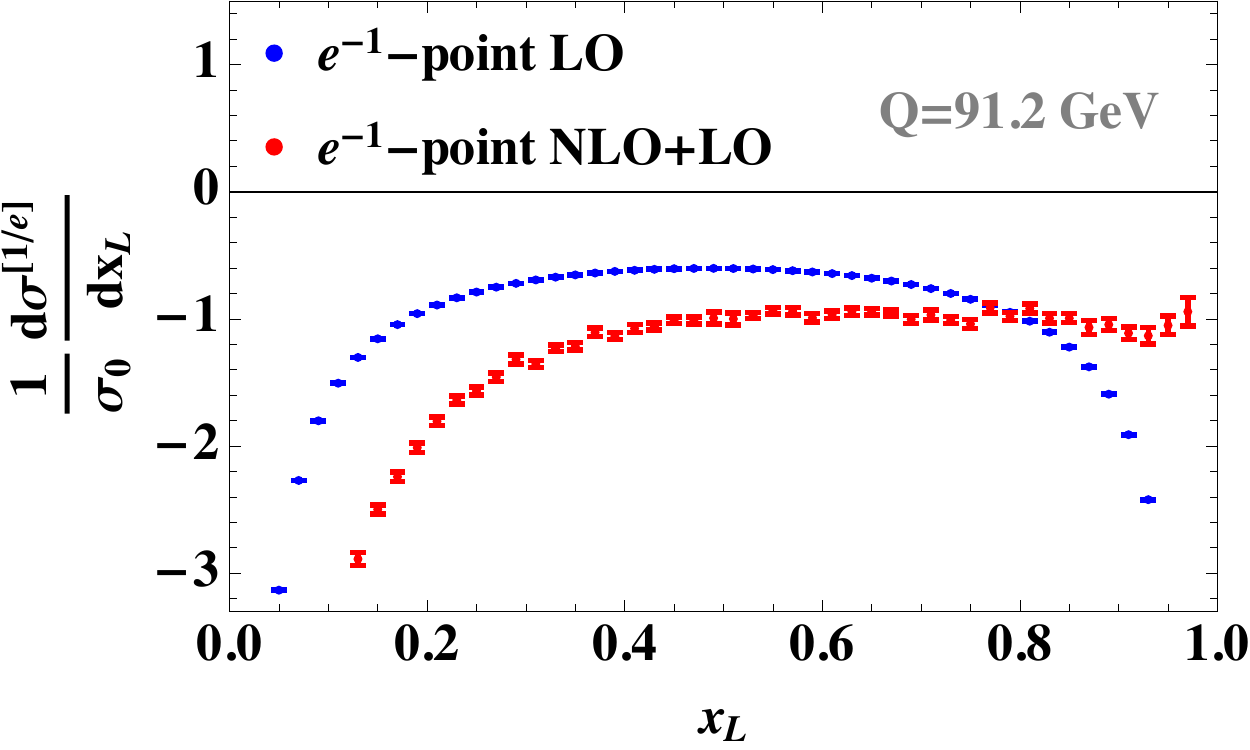}
}
\subfloat[]{\label{fig:inv_e}
\includegraphics[width=0.3\textwidth]{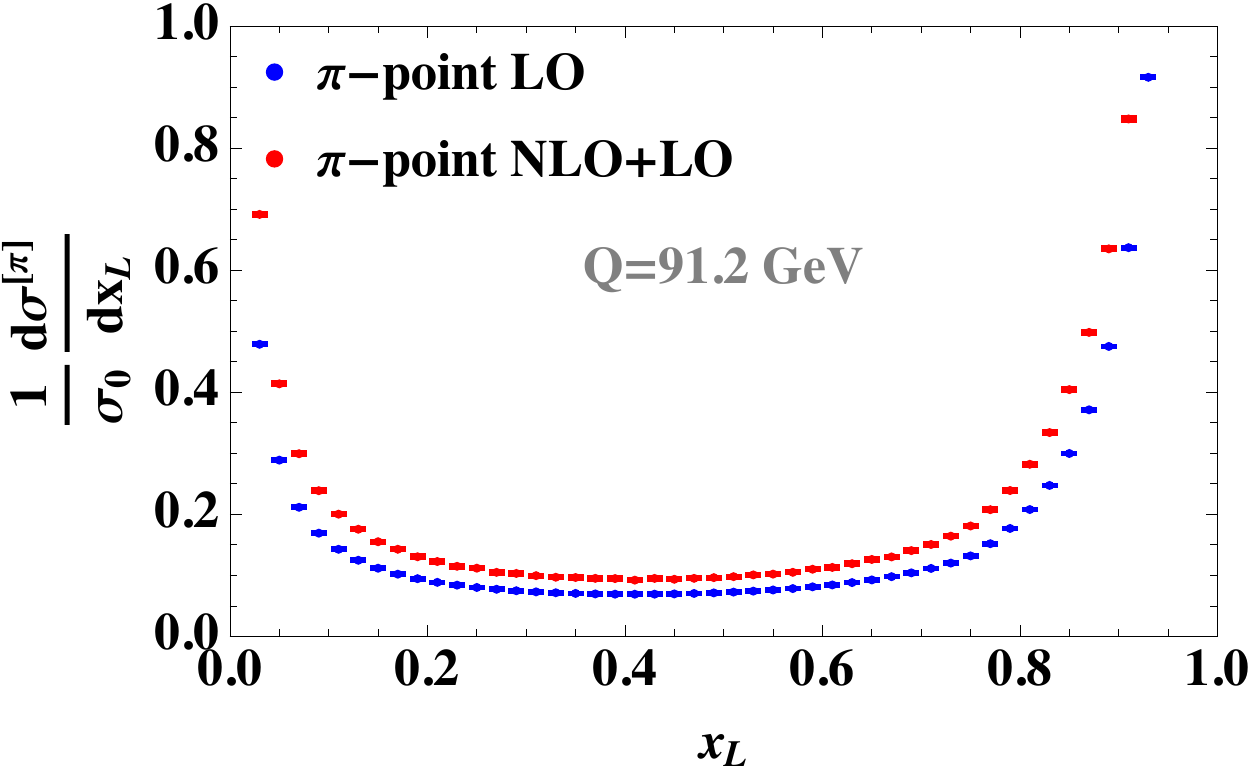}
}
\subfloat[]{\label{fig:complex}
\includegraphics[width=0.3\textwidth]{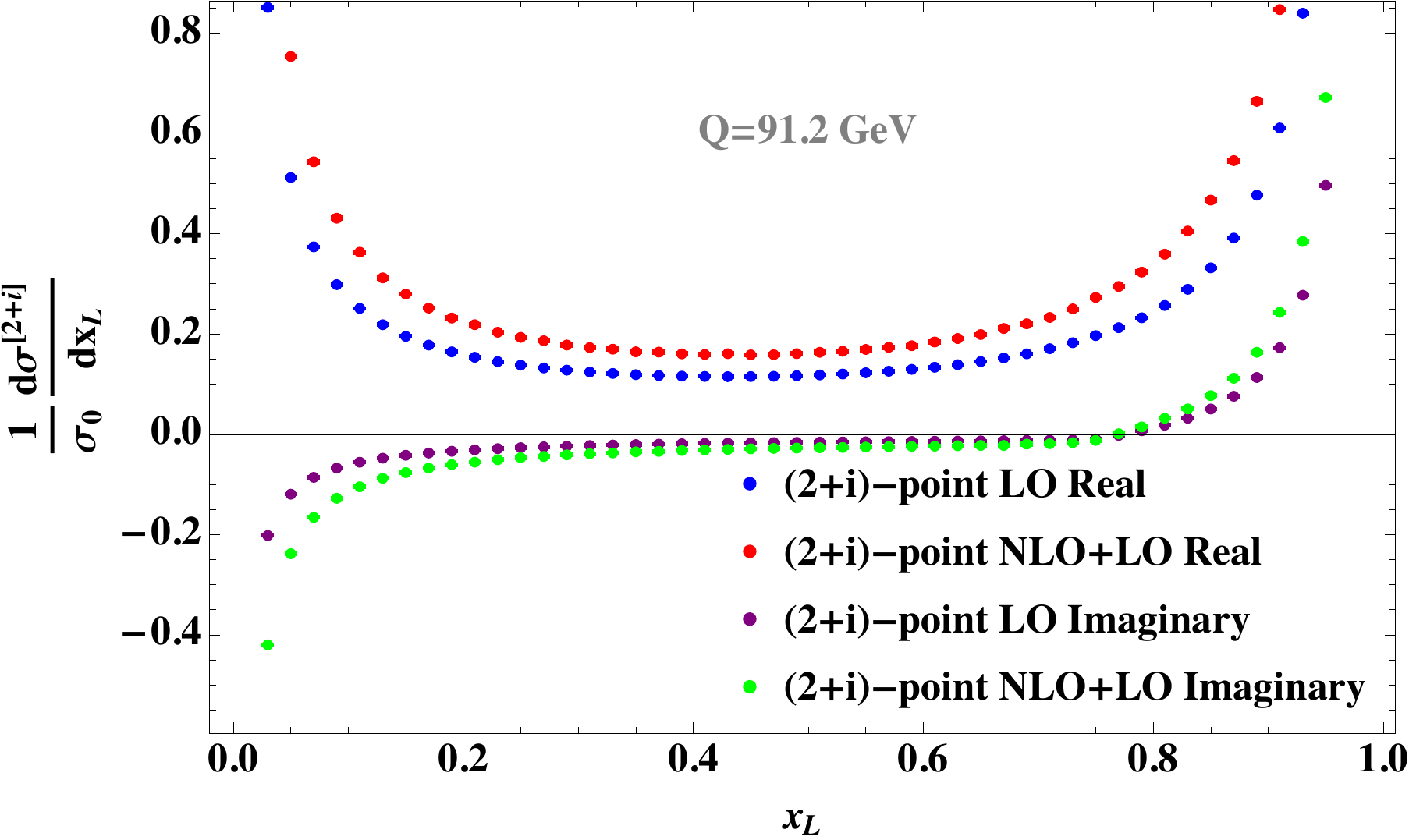}
}
\caption{The LO and NLO predictions for the projected $\nu$-point correlator, for (a) $\nu=e^{-1}$, (b) $\nu=\pi$, and (c) $\nu=2+i$. Calculations were performed numerically with  \texttt{Event2}. Note the qualitatively different behavior for $\nu>1$ and $\nu<1$, as discussed in the text. The finiteness of the distributions for various $\nu$ demonstrate the IRC safety at two loops. Since these are \emph{weighted} cross sections, they are allowed to be negative, or even complex.}\label{fig:event2}
\end{figure*}

\subsection{Ratios of Projected Correlators}
\label{sec:observables_b}

An appealing feature of having multiple observables that depend on the same variable $x_L$, is the ability to take ratios. Such ratios should be more robust experimentally, and are therefore candidates for precision measurements of the strong coupling. 

We define the ratio observable
\begin{align}
\frac{d\sigma^{[N,M]}}{dx_L}= \frac{ \frac{d\sigma^{[N]}}{dx_L} } { \frac{d\sigma^{[M]}}{dx_L}}\,.
\end{align}
The case with $M=2$ and $N=3$ is a particularly interesting candidate observable for precision studies, since it should be calculable with relative ease to NNLL, and we expect that many uncertainties will drop out in the collinear limit. We also expect that hadronization corrections should be minimized for the ratio. Furthermore, the ratio is convenient for probing $\alpha_s$, since as we will show, the variables scale like $d\sigma^{[N]}/dx_L\sim x_L^{\gamma(N+1, \alpha_s)}/x_L$, where schematically $\gamma(N+1,\alpha_s)$ is the Mellin moment of the timelike splitting function. In the ratio the classical $1/x_L$ scaling cancels, and the result is therefore directly proportional to $\alpha_s$, plus higher order corrections. This is analogous to two/three jet ratios that are often used for measurements of the strong coupling, but in the collinear limit within a jet.

\subsection{Higher Point Projections}
\label{sec:observables_c}

We also wish to emphasize that there are numerous other jet substructure observables that can be constructed from the energy correlators to probe increasingly complicated features of jets.  In this paper we have focused on projections to an effective two point correlator.  Beyond two-point correlators, there is no longer just scaling information, but also shape information (and orientation). This shape dependence probes in more detail the structure of the theory. The three point correlator in the collinear limit was computed in \cite{Chen:2019bpb} in $\cN=4$ SYM theory, and in QCD for both quark and gluon jets. It depends on three variables: the longest side $x_L$ and a complex variable $z_3$ defining the position of the third point, as illustrated in \Fig{fig:integration_region}.  It would be interesting to measure the structure of the three point correlator itself, since it provides a detailed probe of the collinear structure of radiation in quark and gluon jets.

Much like how we generalized the two point correlator to the projected $N$-point correlator, we can define a triangle projected $N$-point correlator. For any $N$-points, one can define two triangles by the longest side, and then the third point by the furthest point from either of the two ends of the longest side. Both these triangles are then weighted with the product of the energies of the $N$-partons, much like for the two-point correlator. The two triangles are necessary, since for a triangle one also has an orientation.  We leave the calculation of triangle projected correlators to future work

This construction can of course be done at higher points as well. However, while it is feasible to measure and visualize the four point correlator, it becomes difficult to visualize higher point correlators, since they depend on a large number of variables. Still we believe that it is an interesting question to understand what more general classes of phenomenologically relevant observables can be constructed from the energy correlators.

\section{Analytic Continuation}
\label{sec:analytic}

In this section we discuss a potentially more far reaching consequence of jet (and jet substructure) cross sections expressed directly in terms of energy flow operators, namely that they are amenable to analytic continuation. This will allow us to place all the projected $N$-point correlators into a single analytic family, and to express calculations for arbitrary $N$-point projected correlators in terms of a single analytic function. It will also allow us to define observables that probe $N$-point correlations for non-integer $N$.

Our motivations for this extension are numerous. First, as we will shortly discuss in some detail, for integer values of $N$, the anomalous dimension determining the scaling of the projected $N$-point correlator is the $N$+1 moment of the splitting function. In the case of a conformal field theory, one can further use reciprocity \cite{Mueller:1983js,Dokshitzer:2005bf,Marchesini:2006ax,Basso:2006nk,Dokshitzer:2006nm} to relate this to the $N$+1 moment of the twist-2 spin-$N$+1 anomalous dimensions. In perturbation theory, these are analytic functions of $N$,which has recently been extended to an analytic continuation in spin of the operators \cite{Kravchuk:2018htv}. 
It is therefore interesting to understand if these anomalous dimensions govern the behavior of jet observables. This analytic continuation also provides the potential of directly probing BFKL physics in timelike jets at the LHC, as we will describe in more detail in \Sec{sec:analytic_structure}.

Secondly, analytic continuation of observables places them in a much more rigid structure, which we hope will improve our understanding of jet substructure observables. In the study of jet substructure it is common to speak of observables, such as the angularities \cite{Berger:2003iw}, as a family of observables depending on an angular weight. This angular weighting, which is often called $\beta$, can then be tuned to probe different physics within the jet. However, the angularities are not an analytic function of $\beta$,\footnote{It is possible that with the use of a recoil free axis, angularities could be analytic functions of $\beta$ \cite{Larkoski:2014uqa}.} and the parameter $\beta$ does not have a direct interpretation in the underlying field theory. The $\nu$-correlators introduced in subsection~.\ref{sec:analytic_def} achieve an extension of the two-point correlator that is analogous to the angularities in that there is a single parameter that can be varied, but the $\nu$-correlators are analytic functions of this parameter, and there is a direct operator interpretation of $\nu$ in the field theory.

Finally, and more generally, one would ultimately like more sophisticated ways of designing observables that probe specific field theoretic features of jets. It may turn out that the observables with the simplest analytic properties have complicated algorithmic definitions, potentially involving infinite correlations of particles within jet.  Analytic continuation offers a genuinely new way of constructing jet substructure observables, and may allow for a new organization of observables. Here we will consider observables that probe the twist-2 collinear dynamics of jets, however, one can imagine other analytic families of observables that probe, for example, twist-3 dynamics.

\subsection{Definition of $\nu$-Correlators}
\label{sec:analytic_def}

While it is clear that we can analytically continue analytic functions arising in calculations, what is remarkable is that we are able to present an observable, that can actually be measured on jets of hadrons at the LHC, that corresponds to these analytically continued functions. In this section we will show how this is done, and verify its consistency at next-to-leading order. 

To understand how to analytically continue the $N$-point correlators to generic complex values of $N$, we must think of the observable in a manner that is appropriate for analytic continuation. From here on we will use $\nu$ instead of $N$ to emphasize that we are dealing with non-integer point correlators. The standard $N$-point projected correlator can be understood as measuring the largest angle within each group of $m\leq N$ particles within the jet, and assigning a weight based on the energies of the $m$ particles in the jet. The restriction $m\leq N$ is due to the fact that we can place multiple correlators on the same particle. The relative weightings can be thought of as arising from the binomial formula. For example for the three point correlator with three particles we have, 
\begin{align}
&(E_1+E_2+E_3)^3=6(E_1E_2E_3) 
+ (E_1^3+E_2^3+E_3^3) \\
&\hspace{-0.3cm}+(3E_1^2 E_2+3E_1^2 E_3+3 E_2^2 E_3+3 E_1^2 E_1 +3 E_3^2 E_1+3 E_3^2 E_2)\,, \nn
\end{align}
where the first line describes energy correlators placed on three distinct particles, and the second and third line describe contact terms. Observables constructed in this manner are guaranteed to be infrared and collinear safe, at least when integrated over the angles, due to the sum rule in \eqn{eq:sumrule_N}. Since these are polynomial weightings, they provide a starting point for performing the analytic continuation. It should be intuitively clear at this point that for generic values of $\nu$, the analytically continued observables will probe infinite correlations, since the expansion of $(E_1+E_2+E_3)^\nu$ using the binomial theorem does not collapse to a finite sum unless $\nu$ is an integer.

We define the analytic continuation of the $N$-point correlator, which we call a $\nu$-correlator, as
\begin{widetext}
\begin{align}
  \label{eq:projected_mom_nu}
 \frac{d \sigma^{[\nu]}}{dx_L} &\ = \sum_n \int \! d\sigma_{e^+e^- \to X_n} 
 \cdot 
\Big[
 \sum_{1\leq i_1 \leq n} \cW_1^{[\nu]}(i_1) \delta(x_L)  +
\sum_{1 \leq i_1 < i_2 \leq n} \cW_2^{[\nu]} (i_1, i_2) \delta(x_L - x_{i_1 i_2} )
\nn\\
&\ + 
\sum_{1 \leq i_1 < i_2 < i_3 \leq n} \cW_3^{[\nu]} (i_1, i_2, i_3) \delta(x_L - \max \{x_{i_1 i_2} , x_{i_1 i_3}, x_{i_2 i_3} \} ) + \ldots
\nn\\
&\ + 
\sum_{1 = i_1 < i_2 < \ldots < i_n = n} \cW_n^{[\nu]} (i_1, i_2, \ldots, i_n) \delta(x_L - \max \{x_{i_1 i_2} , x_{i_1 i_3}, \ldots, x_{i_{n-1} i_n}  \} ) \Big]  \,,
\end{align}
where each term in the square bracket probe a specific number of particles being measured.
In the last line the summation collapses into a single term. The weights for the different numbers of particles being correlated are
\begin{align}
  \label{eq:weight_func}
  \cW_1^{[\nu]} (i_1) = &\ \frac{E_{i_1}^\nu}{Q^\nu} \,,
\nn\\
  \cW_2^{[\nu]} (i_1, i_2) = &\ \frac{(E_{i_1} + E_{i_2})^\nu}{Q^\nu} - \sum_{1 \leq a \leq 2} \cW_1^{[\nu]}(i_a) \,,
\nn\\
\cW_3^{[\nu]}(i_1, i_2, i_3) = &\ \frac{(\sum_{a=1}^3 E_{i_a})^\nu}{Q^\nu} - \sum_{1 \leq a < b \leq 3} \cW_2^{[\nu]}(i_a, i_b) - \sum_{1 \leq a \leq 3} \cW_1^{[\nu]}(i_a) \,,
\nn\\
\ldots &\  \,,
\nn\\
  \cW_n^{[\nu]} (i_1, \ldots, i_n) = &\ \frac{\left( \sum_{a=1}^n E_{i_a}\right)^\nu}{Q^\nu} 
- \sum_{1 \leq a_1 < a_2 < \ldots < a_{n-1} \leq n} \cW_{n-1}^{[\nu]}(i_{a_1} , i_{a_2} , \ldots , i_{a_{n-1}})
- \ldots - \sum_{1 \leq a \leq n }^n \cW_1^{[\nu]}(i_a) \,.
\end{align}
\end{widetext}
This observable obeys, by construction, the sum rule
\begin{align}
  \label{eq:sumrule_nu}
  \int_0^1 \! dx_L \, \frac{d\sigma^{[\nu]}}{dx_L} = \sigma_{\rm tot} \,.
\end{align}
As expected, for generic values of $\nu$ this observable involves correlations of an infinite number of particles. However, for integer values of $\nu$, the sum collapses. Taking $\nu=N$, one can easily check that  $\cW_n^{[N]}=0$ for $n>N$.
In reality, the sum also collapses because there are only finitely many particles in a collision event. It is therefore realistic to measure experimentally, although an efficient algorithm will need to be developed when the particle number is large. As with the case of the projected $N$-point correlator, the $\nu$-correlator can be straightforwardly defined on jets by restricting the sum to particles within the jet. 

\subsection{Infrared Safety at Fixed Order}
\label{sec:analytic_calc}

We now give a proof that Eq.~\eqref{eq:projected_mom_nu} is IRC safe at the first non-trivial order in perturbation theory. We use $e^+ e^- \to q \bar q$ as an example, and work to $\cO(\alpha_s)$, to illustrate a non-trivial soft and collinear cancellation. The well-known KLN theorem~\cite{1964PhRv..133.1549L,1962JMP.....3..650K} states that inclusive cross sections in $e^+e^-$ are infrared finite to all orders in perturbation theory. At $\cO(\alpha_s)$, the inclusive cross section can be separated into one-loop two-particle final states~(virtual corrections) and tree-level three-particle final states~(real corrections). While their individual contributions diverge, their sum after integration over their respective phase spaces is finite,
\begin{align}
  \label{eq:KLN}
  \int\! d \sigma_{V, q\bar q} +   \int\! d \sigma_{R, q\bar q g} = \frac{\alpha_s}{\pi} \sigma_0 \,.
\end{align}
In particular, $d\sigma_{V, q\bar{q}} = {\cal V}(\alpha_s, \e) d\sigma_0$ contains explicit IR poles,
\begin{align}
  \label{eq:virt_poles}
 {\cal V}(\alpha_s, \e) = \frac{\alpha_s}{4 \pi} C_F \left( - \frac{4}{\e^2} - \frac{6}{\e} + \text{finite terms} \right)  \,,
\end{align}
where we have set $\mu = Q$ for simplicity. On the other hand, the differential three-body cross section $d \sigma_{R, q\bar{q} g}$ is finite. Divergences arise only after integration over phase space. 
We shall consider virtual and real corrections separately. 

For the virtual corrections, we have
\begin{align}
  \label{eq:virt_corr}
  \frac{d \sigma_V^{[\nu]}}{d x_L} &\ = 
\int \! d\sigma_{V, q\bar q} \Big[ 
\Big( \cW_1^\supnu(1) + \cW_1^\supnu(2) \Big) \delta(x_L)
\nn\\
&\ 
+ \cW_2^\supnu (1, 2) \delta(x_L - 1) 
\Big]  \,,
\end{align}
where in Eq.~\eqref{eq:virt_corr} the weight functions are given by
\begin{gather}
  \label{eq:virt_weight_func}
  \cW_1^\supnu (1)\Big|_{q \bar q} =  \cW_1^\supnu (2)\Big|_{q \bar q} = 2^{-\nu} \,,
\\
 \cW_2^\supnu (1, 2)\Big|_{q \bar q} =   1 - 2^{1 - \nu} \,.
\end{gather}
The weight function will in general be different for different numbers of particles in the final state, for which we use a subscript to denote. Virtual corrections contribute only to the end point of the $\nu$-correlator.

The real corrections can be written as
\begin{align}
  \label{eq:real_corr}
    \frac{d \sigma_R^{[\nu]}}{d x_L} &\ = 
\int \! d\sigma_{R, q\bar q g} \Big[ 
\Big( \cW_1^\supnu(1) + \cW_1^\supnu(2) + \cW_1^\supnu(3) \Big) \delta(x_L)
\nn\\
&\ 
+ \Big( \cW_2^\supnu (1, 2) \delta(x_L - x_{12}) + \cW_2^\supnu (1, 3) \delta(x_L - x_{13}) 
\nn\\
&\
+ \cW_2^\supnu (2, 3) \delta(x_L - x_{23}) \Big) 
\nn\\
&\
+
\cW_3^\supnu (1, 2, 3) \delta(x_L - \max\{x_{12}, x_{13}, x_{23} \})
\Big]  \,.
\end{align}
We divide the three-body phase space into hard, $qg$ collinear, $\bar{q}g$ collinear, and large-angle soft radiation region according to the infrared behavior of QCD matrix element. In the hard region, the final states are non-degenerate, and $x_{12}$, $x_{13}$, $x_{23}$ take generic value between $0$ and $1$. This region is clearly IRC finite. 

We now consider the $qg$ collinear limit, $1 \parallel 3$. In this region, we have $x_{13} = 0$, $x_{12} = x_{23} = 1$. The real corrections in this region become
\begin{align}
  \label{eq:real_13}
 &\   \frac{d \sigma_{R,1\parallel 3}^{[\nu]}}{d x_L}  = 
\int_{1 \parallel 3} \! d\sigma_{R, q\bar q g} \Big[ 
\Big( \cW_1^\supnu(1) + \cW_1^\supnu(2) + \cW_1^\supnu(3)
\nn\\
&\  + \cW_2^\supnu(1,3) \Big) \delta(x_L)
\nn\\
&\ 
+ \Big( \cW_2^\supnu (1, 2) 
+ \cW_2^\supnu (2, 3)
 + \cW_3^\supnu (1, 2, 3) \Big) \delta(x_L - 1)
\Big]  \,.  
\end{align}
Using Eq.~\eqref{eq:weight_func}, we can simplify this to
\begin{align}
  \label{eq:real_13_a}
 &\   \frac{d \sigma_{R,1\parallel 3}^{[\nu]}}{d x_L}  = 
\int_{1 \parallel 3} \! d\sigma_{R, q\bar q g} \Big[ 
\Big( \cW_1^\supnu(2) + \frac{(E_1 + E_3)^\nu}{Q^\nu}   \Big) \delta(x_L)
\nn\\
&\ 
+ \Big( \frac{(E_1 + E_2 + E_3)^\nu}{Q^\nu} - \frac{(E_1+E_3)^\nu}{Q^\nu} - \cW_1(2)  \Big) \delta(x_L - 1)
\nn\\
&\ =
\int_{1 \parallel 3} \! d\sigma_{R, q\bar q g} \Big[ 
2^{1 - \nu} \delta(x_L) + (1 - 2^{1 - \nu}) \delta(x_L - 1) 
\Big]  \,, 
\end{align}
where in the second equality we have used the collinear kinematics, $E_1 + E_3 = Q/2$. The $\bar{q} g$ collinear limit is identical due to charge conjugate invariance of QCD,
\begin{align}
  \label{eq:eq:real_23}
   \frac{d \sigma_{R,2\parallel 3}^{[\nu]}}{d x_L}  =  \frac{d \sigma_{R,1\parallel 3}^{[\nu]}}{d x_L}   \,.
\end{align}

We now consider the large-angle soft radiation region, $3_s$. We have $x_{12} = 1$, and $x_{13}$ and $x_{23}$ take generic values between $0$ and $1$. The real corrections become
\begin{align}
  \label{eq:real_soft}
   &\   \frac{d \sigma_{R,3_s}^{[\nu]}}{d x_L}  = 
\int_{3_s} \! d\sigma_{R, q\bar q g} \Big[ 
\Big( \cW_1^\supnu(1) + \cW_1^\supnu(2) + \cW_1^\supnu(3) \Big) \delta(x_L)
\nn\\
&\ +
\Big(\cW_2^\supnu(1,2) + \cW_3^\supnu(1,2,3) \Big) \delta(x_L - 1)
\nn\\
&\
+ \cW_2^\supnu(1,3) \delta(x_L - x_{13})
+ \cW_2^\supnu(2,3) \delta(x_L - x_{23})
\Big]
\nn\\
&\ = \int_{3_s} \! d\sigma_{R, q\bar q g} \Big[ 
(2^{1 - \nu} + \cW_1^\supnu(3)) \delta(x_L)
\nn\\
&\
+ (1 - 2 ^{1 - \nu} + \cW_1^\supnu(3)) \delta(x_L - 1)
\nn\\
&\
- \cW_1^\supnu(3) \delta(x_L - x_{13}) 
- \cW_1^\supnu(3) \delta(x_L - x_{23}) 
\Big] \,,
\end{align}
where
\begin{align}
\cW_1^\supnu(3) = \frac{E_3^\nu}{Q^\nu} \,,
\end{align}
vanishes for $\Re (\nu) > 0$. We have shown the IR singularities reside in the end point in the individual contributions. Adding the different regions together, we find that
\begin{align}
  \label{eq:virt_real_sing}
&\  \frac{d\sigma_V^\supnu}{d x_L} + \frac{d\sigma_{R,1\parallel 3}^\supnu}{d x_L}
 + \frac{d\sigma_{R,2\parallel 3}^\supnu}{d x_L}
+ \frac{d\sigma_{R,3_s}^\supnu}{d x_L}
\\
\stackrel{\Re(\nu)>0}{=} &\ [  2^{1 - \nu}\delta(x_L) 
+ ( 1 - 2^{1 - \nu} ) \delta(x_L - 1) ]
\nn\\
&\
\cdot \left[ 
\int\! d\sigma_{R, q \bar q} + 
\left(\int_{1 \parallel 3} \! +
\int_{2 \parallel 3} \! +
\int_{3_s} \! \right)
d\sigma_{R, q \bar q g}
\right] \,. \nn
\end{align}
The third line is IRC  finite by the KLN theorem. We have therefore shown that the projected $\nu$-point correlator is IRC safe at this order. For $\nu = 1$, there is no back-to-back end-point contribution, $\delta(x_L - 1)$. This agrees with the expectation that $\nu = 1$ corresponds to the $1$-point correlator, which only has collinear end-point contribution, $\delta(x_L)$.

We have therefore shown that the $\nu$-correlator  is IRC safe at $\Ord(\alpha_s)$ for $\nu > 0$. As $\nu$ decreases, the $\nu$-correlator  is increasingly sensitive to low energy soft gluon radiation. It therefore also provides a probe of non-perturbative soft physics. Since $\nu$ is a tunable parameter, the $\nu$-correlators provide a convenient way to experimentally probe different aspects of QCD dynamics in a single style of measurement.

\begin{figure*}[t]
\centering
\subfloat[]{\label{fig:chew_frautschi}
\includegraphics[width=0.4\textwidth]{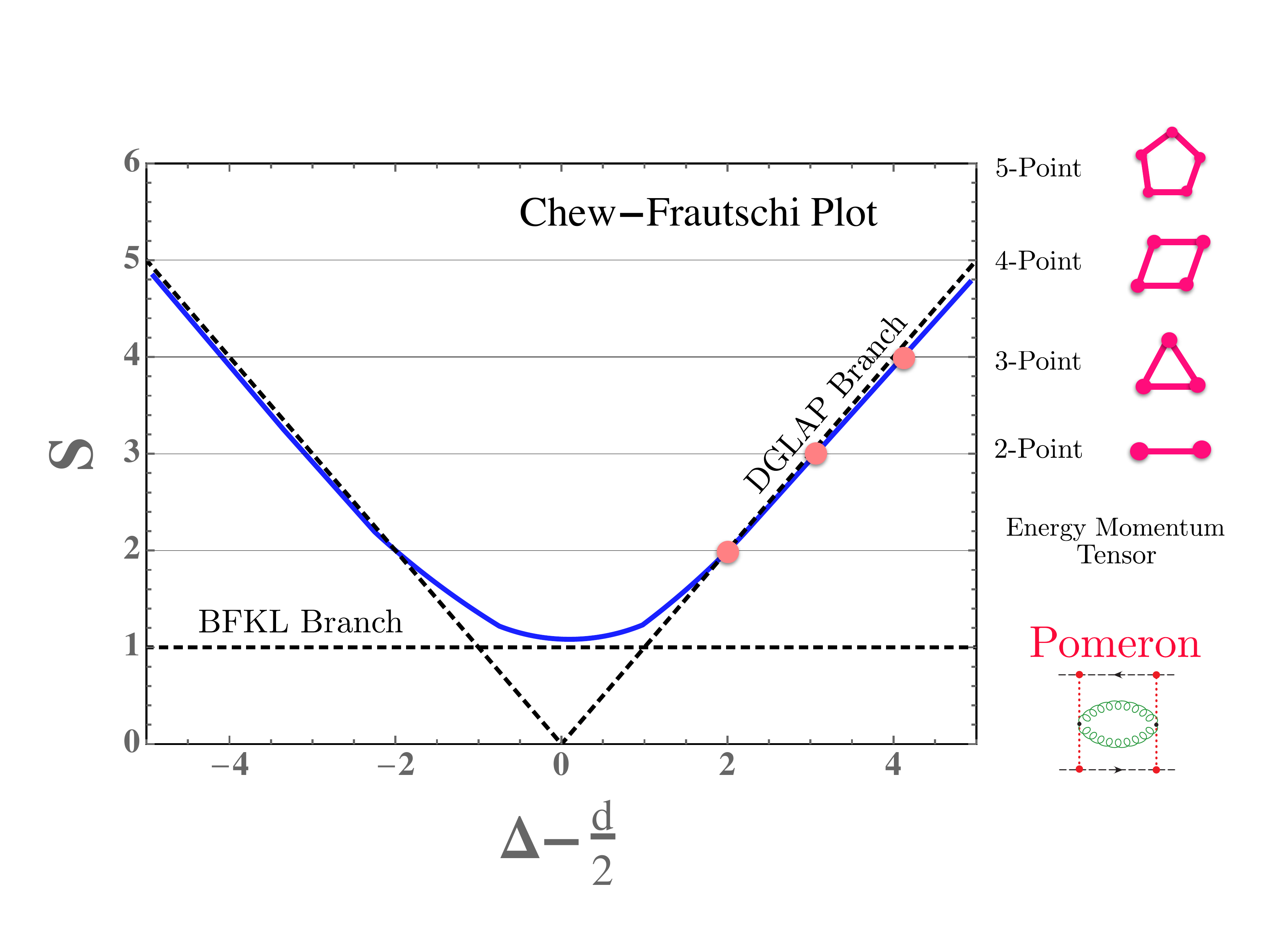}
}
\subfloat[]{\label{fig:nu_plane}
\includegraphics[width=0.4\textwidth]{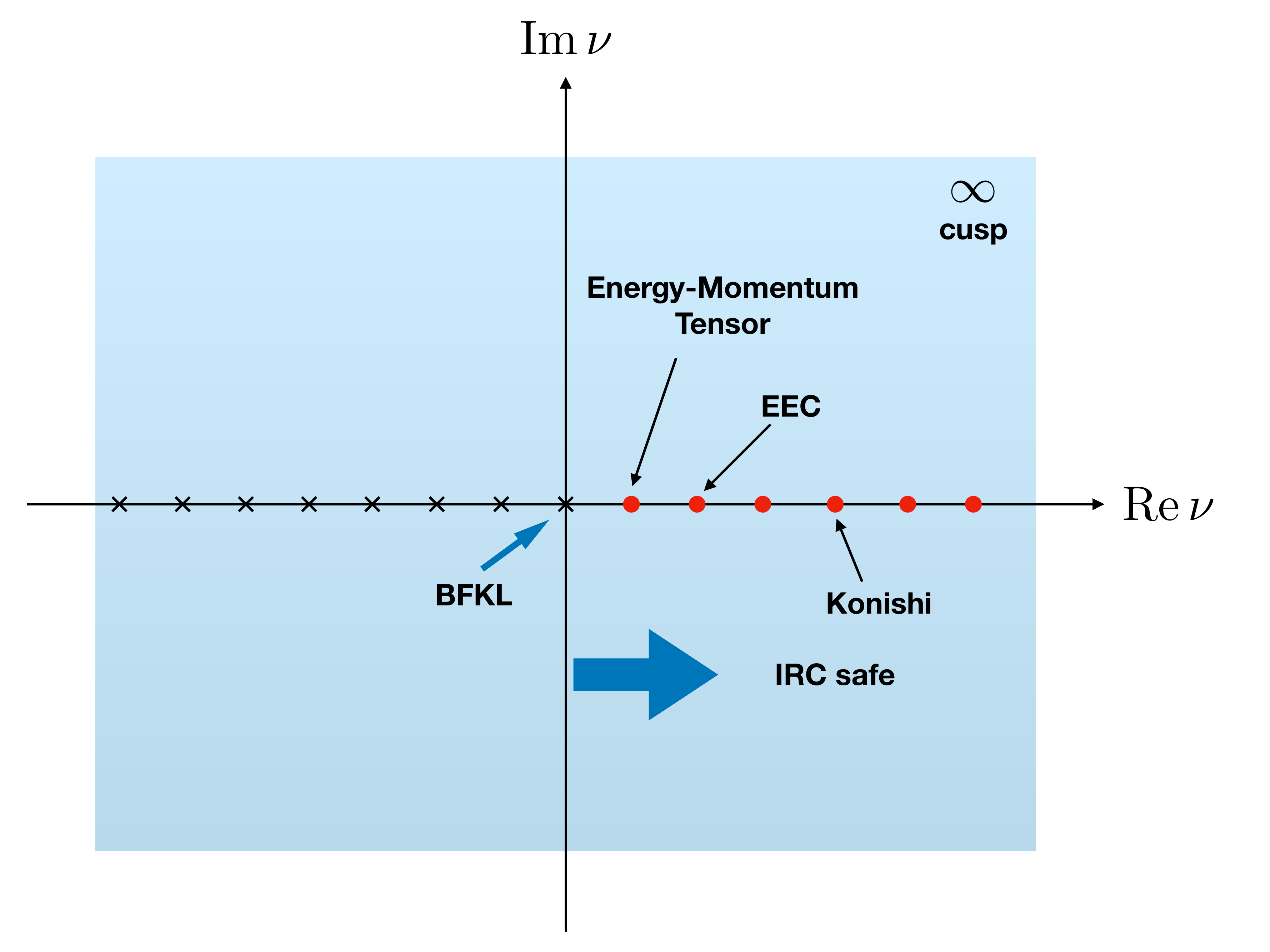}
}
\caption{(a) The analytically continued $\nu$ point correlators probe the analytic family of twist-2 spin-j operators using the collinear physics of jets. For integer values of $\nu$ these collapse to the standard $N$-point correlators. (b) In the complex $\nu$-plane the observable exhibits poles at negative integer values, related to BFKL physics. The pole at $\nu=0$, corresponding to the BFKL pomeron in the spacelike case, is associated with multiplicity in the time like case. As $\nu\to \infty$, one observes logarithmic growth in $\nu$ associated with the cusp anomalous dimension. }\label{fig:analytic}
\end{figure*}

\subsection{Analytic Structure of the $\nu$-Plane}
\label{sec:analytic_structure}

We now discuss in more detail the physics of the $\nu$-correlators. As we shall show in Sec.~\ref{sec:N}, the scale evolution of the $\nu$-correlators with $x_L$ is determined by the twist-2 spin-$\nu+1$ anomalous dimensions up to running coupling effects. The twist-2 anomalous dimensions are well known to have a rich analytic structure, for example, enabling analytic continuation between the DGLAP and BFKL regimes \cite{Jaroszewicz:1982gr,Lipatov:1996ts,Kotikov:2000pm,Kotikov:2002ab,Kotikov:2004er,Brower:2006ea} (For a detailed review of the analytic properties of the twist-2 anomalous dimensions, see \cite{Kotikov:2019xyg}.). More recently, there has been renewed interested in the analytic properties of these operators in the context of conformal field theories \cite{Caron-Huot:2017vep,Kravchuk:2018htv}. One can therefore hope that this analytic structure is reflected in the behavior of the $\nu$-correlators, which can be measured in collider experiments. Here we highlight some of the key features  of the $\nu$-correlators in the $\nu$-plane in \Fig{fig:nu_plane}. The resummation of the $\nu$-correlators for generic values of $\nu$ will be presented in \Sec{sec:N}, and will provide additional insight.

For positive integer $\nu$, the $\nu$-correlators correspond to standard $N$-point correlators which evolve with the anomalous dimensions of the twist-2 spin-$N+1$ operators, which are standard local operators. The case $N=2$ has received the most attention \cite{Dixon:2019uzg,Korchemsky:2019nzm,Kologlu:2019mfz}. Another positive integer value of particular interest is $\nu=1$. In terms of the matrix elements of energy flow operators, this corresponds to a three-point function, which in a CFT is completely fixed by symmetry. This has been discussed in detail in \cite{Hofman:2008ar}. In the context of QCD, the $\nu=1$ case is the well-known semi-inclusive hadron production in $e^+e^-$, where perturbative coefficients have been computed to NNLO~\cite{Mitov:2006ic,Almasy:2011eq}. More generally, since the anomalous dimension of the stress tensor vanishes, the $\nu=1$ point correlator will not exhibit any non-trivial scaling behavior. For $\nu=4$, the relevant scaling anomalous dimension corresponds to that of the Konishi operator \cite{Konishi:1983hf}. Finally, as $\nu\to \infty$, the twist-2 anomalous dimensions scale like $\gamma(j)\propto \Gamma_{\text{cusp}} \ln(j)$ as $j\to \infty$ (Here we use $\nu = j - 1$ as is common.), where $\Gamma_{\text{cusp}}$ is the cusp anomalous dimension \cite{Polyakov:1980ca,Korchemsky:1987wg}. Its physical appearance here follows from arguments analogous to those presented in \cite{Korchemsky:1988si}. It is also important to comment on the region of applicability of our result as $\nu\to \infty$. For the results given in this section we have worked at leading twist. However, even at weak coupling, due to the logarithmic growth of the twist-2 anomalous dimension, at sufficiently large $j$, one has a level crossing with the twist-four operators. This level crossing has been studied explicitly in \cite{Korchemsky:2015cyx}. While it is theoretically interesting, it occurs when $j\simeq e^{\frac{\pi}{\alpha_s N_c}}\,,$ which in perturbation theory seems to be well beyond what could be considered practically in experiments.

A phenomenologically interesting region is the analytic continuation towards $\Re (\nu) =0$. It is well known that both the spacelike and timelike anomalous dimensions diverge in this limit in fixed order perturbation theory. At lowest order, this behavior is a power law $\gamma(\nu) \propto 1/\nu$. The anomalous dimension itself must be resummed to have a well defined scaling in this limit. Although the EEC is naively a timelike measurement, one can use reciprocity~\cite{Basso:2006nk,Marchesini:2006ax} to show that the scaling of the observable is determined by the spacelike anomalous dimension in a conformal field theory~\cite{Dixon:2019uzg,Korchemsky:2019nzm}. When conformal symmetry is broken, it naively seems like it must be formulated as a timelike problem. 

In the case of a CFT, reciprocity allows the behavior as $\nu\to 0$ to be interpreted in terms of the BFKL pomeron. The fact that BFKL dynamics can appear in jet physics \cite{Hatta:2008st} may be surprising, but arises due to a conformal mapping relating the transverse plane in BFKL dynamics to the celestial sphere in $e^+e^-$ annihilation \cite{Hatta:2008st,Caron-Huot:2015bja,Caron-Huot:2016tzz} (for a recent discussion see \cite{Neill:2020bwv}). 
 The BFKL theory \cite{Kuraev:1977fs,Balitsky:1978ic,Lipatov:1985uk} describes the behavior of the twist-2 anomalous dimensions in this limit. In particular, the BFKL equation \cite{Kotikov:2007cy}
\begin{align}
\frac{\nu}{-4g^2}=\Psi\left(  -\frac{\gamma}{2} \right)+\Psi\left(  1+\frac{\gamma}{2} \right)-2\Psi(1)\,,
\end{align}
relates the anomalous dimension $\gamma$ and $\nu$ in this limit. Inverting this equation, we have the behavior of $\gamma$ as $\nu\to 0$ \cite{Kotikov:2007cy}
\begin{align}
\gamma=2 \left( \frac{-4g^2}{\nu}  \right) -4\zeta_3 \left(  \frac{-4g^2}{\nu}    \right)^4 +\cdots\,.
\end{align}
Therefore the divergence of the anomalous dimension which controls the scaling of the measurable jet observable is controlled by the BFKL equation.

In the timelike case, the resummation of the anomalous dimension as $\nu\to 0$ has been studied in the context of multiplicity. There it is well known (see e.g. \cite{Ellis:1991qj,Ioffe:2010zz}) that as $\nu\to 0$, the anomalous dimension takes the form
\begin{align}
\gamma = -\frac{1}{4} \left( \sqrt{\nu^2+\frac{8C_A \alpha_s}{\pi}} -\nu   \right)\,,
\end{align}
which now has a finite limit as $\nu\to 0$. As we will discuss in more detail shortly, this provides some insight into the physical interpretation of the $\nu$-correlator as $\nu\to 0$ as a form of multiplicity correlation. Multiplicity itself is both soft and collinear unsafe, but can be made soft safe after resummation of the scaling anomalous dimension. Here $\nu$, which corresponds to the energy weighting in the observable is tracking the soft safety of the observable, while the resolution parameter $x_L$, is tracking the collinear safety. We will see later that as $x_L\to 0$, the $\nu$ point correlator for $\nu<1$ diverges, corresponding to the fact that multiplicity and multiplicity fluctuations diverge as the scale at which they are probed (loosely the infrared regulator) is taken to zero.

There is also an intriguing parallel with the analytic continuation of twist-2 operators considered in \cite{Balitsky:1987bk,Kravchuk:2018htv}. There the operators which analytically continue the twist-2 operators to non-integer spin collapse to local operators at integer values. This is analogous to how the $\nu$-correlators collapse to correlating finite numbers of particles within jets for integer values of $\nu$. It would be interesting to understand more formally this connection.

\section{Resummation for the  $\nu$-Correlator}
\label{sec:N}

In this section we discuss the factorization and resummation of the $\nu$-correlator for generic values of $\nu$, and present results through NLL accuracy. The resummed results also provide considerable insight into the physical interpretation of the $\nu$-correlators for non-integer values of $\nu$ that was discussed in the previous section.

\subsection{Factorization Formula}
\label{sec:fact-form}

In the small angle limit, we propose a timelike factorization formula for the  $\nu$-correlator
\begin{align}
  \label{eq:fac_nu}
  \Sigma^\supnu(x_L, \ln\frac{Q^2}{\mu^2}) = \int_0^1 \! dx\, 
x^\nu \vec{J}\,^\supnu(\ln \frac{x_L x^2 Q^2}{\mu^2}) \cdot \vec{H}(x, \frac{Q^2}{\mu^2}) \,,
\end{align}
where we have suppressed the $\alpha_s(\mu)$ dependence in all functions. This is an extension of the factorization formula for the EEC presented in \cite{Dixon:2019uzg}. This factorization holds both in conformal and non-conformal theories.  The hard function satisfies the timelike DGLAP evolution equation,
\begin{align}
  \label{eq:hard_evo}
  \frac{d \vec{H}(x, \ln\frac{Q^2}{\mu^2})}{d \ln \mu^2}
= - \int_x^1\! \frac{dy}{y} \widehat{P}(y) \cdot \vec{H} \left( \frac{x}{y},
\ln\frac{Q^2}{\mu^2}\right) \,,
\end{align}
where $\widehat{P}(y,\alpha_s)$ is the singlet timelike splitting matrix. 
From RG invariance of the physical cross section, we find that the jet function satisfies a modified timelike DGLAP evolution equation,
\begin{align}
  \label{eq:jet_evo}
  \frac{d \vec{J}\,^\supnu( \ln \frac{x_L Q^2}{\mu^2})}{d \ln\mu^2}
=
\int_0^1 \! dy\, y^\nu \vec{J}\,^\supnu( \ln \frac{x_L y^2 Q^2}{\mu^2})
\cdot 
\widehat{P}(y) \,.
\end{align}
This is one of the main results in this paper. It is surprising that while the measurement defined by the $\nu$-correlator can become quite involved, its scale dependence is simple and is fixed completely by RG invariance argument. This illustrates the power of factorization. 

\subsection{Hard Function}
\label{sec:hard-function}

The hard functions for the $\nu$-correlators are equal to the coefficient functions for semi-inclusive hadron fragmentation~\cite{Mitov:2006ic,Almasy:2011eq}, and are only sensitive to the hard scale $Q$ of the problem. They are vectors in flavor space,
\begin{align}
  \label{eq:hard_func}
  \vec{H}(x, \ln \frac{Q^2}{\mu^2}) = (H_q(x,\ln \frac{Q^2}{\mu^2}), H_g(x,\ln \frac{Q^2}{\mu^2}))  \,,
\end{align}
where $H_q(x, \ln Q^2/\mu^2)$ is the probability of finding a quark~(or anti-quark) with momentum fraction $x = (2 p \cdot q)/Q^2$, where $p$ is the momentum of the quark, and $q^2 = Q^2$, and similarly for $H_g$. We consider two processes in this paper: $e^+e^-$ annihilation, $\vec{H}^{\rm ee}$, and Higgs decay, $\vec{H}^{\rm h}$. To achieve the NLL accuracy considered in this paper, we need the hard functions to NLO, which we give in \App{app:hard_func}.

\subsection{Jet Function}
\label{sec:jet-function}

The jet function, which depends on the details of the measurement (and hence $\nu$) is a vector in flavor space,
\begin{align}
\vec{J}\,^\supnu = \begin{pmatrix}
J_q^\supnu
\\
J_g^\supnu 
\end{pmatrix} \,.
\end{align}
We expand the jet function in the strong coupling constant as
\begin{align}
J_q^\supnu = J_0^{q, [\nu]} + \frac{\alpha_s}{4 \pi} J_1^{q, [\nu]} + \left( \frac{\alpha_s}{4 \pi} \right)^2 J_2^{q, [\nu]} + \ldots \,,
\end{align}
and similarly for the gluon jet function. The LO jet function is given by
\begin{equation}
  \label{eq:jet_LO}
  J_0^{q, [\nu]} = J_0^{g, [\nu]} = 2^{-\nu} \,.
\end{equation}
We have chosen a slightly different normalization for the jet function as compared with Ref.~\cite{Dixon:2019uzg}. The $2^{-\nu}$ factor arises because here we normalize the energy correlators to $Q^{-\nu}$, which at LO is twice the jet $p_T$. If instead  we normalized the energy correlators to $(p_T^\text{jet})^{-\nu}$, the overall factor of $2^{-\nu}$ would be absent. The latter normalization may be convenient for jet production at the LHC. We keep track of the $2^{-\nu}$ factor when analytic formulas are presented, so that conversions between the different normalizations is straightforward.

 At one loop order the jet function can be calculated from the QCD $1 \to 2$ timelike splitting kernel, 
 \begin{align}
   \label{eq:jet_one_loop}
   \frac{\alpha_s}{4 \pi} J_1^{i, [\nu]} = &\
 \frac{\mu^{2 \e} e^{\e \gamma_E}}{(4 \pi)^\e}  
\int_0^1 \! dx \int_0^{x_L x (1-x) Q^2 } \! ds  \frac{[ x (1-x) s]^{- \e}}{ (4 \pi)^{2 - \e} \Gamma(1-\e) }
\nn\\
&\
\cdot
\frac{2 g^2}{s} P_{i \to 12} (x) \cW_2^\supnu (1, 2) \,,
 \end{align}
where $\gamma_E  = 0.577216 \ldots$ is Euler's gamma constant, $s$ is the invariant mass of the splitting pair, and $x$ is the momentum fraction of the daughter particle $1$, and
\begin{align}
W_2^\supnu(1,2) = 2^{- \nu} ( 1 - x^\nu - (1 - x)^\nu) \,.
\end{align}
To this order, the relevant fragmentation kernels are
\begin{align}
  \label{eq:frag_kernel}
  P_{gq}(x) =&\ C_F \left( \frac{1 + (1 - x)^2}{x} - \e x \right)\,,
\nn\\
P_{gg}(x) = &\ 2 C_A \frac{ ( 1 - x + x^2)^2}{x (1-x) } \,,
\nn\\
P_{qg}(x) = &\ \frac{1}{2} \left( 1 - \frac{2 x (1-x)}{1- \e} \right) \,.
\end{align}
We find the bare one-loop jet function to be~(for the $g \to gg$ splitting, an additional symmetry factor $1/2$ is needed in the phase space of  Eq.~\eqref{eq:jet_one_loop})
\begin{widetext}
  \begin{align}
    \label{eq:jet_cst_one_loop}
   2^\nu J_1^{q, [\nu]} = &\ C_F \Bigg[ \frac{3 (\nu -1)-4 (\nu +1) (\Psi(\nu) + \gamma_E)}{\nu +1} \left(\frac{1}{\e} - \ln \frac{x_L Q^2}{\mu^2} \right) 
\nn\\
&\
+
\frac{13 \nu ^3+24 \nu ^2-25 \nu -12}{\nu  (\nu +1)^2}-4 (\Psi(\nu) + \gamma_E)^2-\frac{12
   (\Psi(\nu) + \gamma_E)}{\nu +1}+12 \Psi'(\nu)-2 \pi ^2 \Bigg]
+ \cO(\e) \,,
\nn\\
2^\nu J_1^{g, [\nu]} = &\
\left[ C_A \left(\frac{(\nu -1) \left(11 \nu ^2+53 \nu +66\right)}{3 (\nu +1) (\nu +2) (\nu
   +3)}-4 (\Psi(\nu) + \gamma_E)\right)-\frac{2 (\nu -1) \left(\nu ^2+4 \nu +6\right) n_f}{3 (\nu
   +1) (\nu +2) (\nu +3)} \right] \left(\frac{1}{\e} - \ln \frac{x_L Q^2}{\mu^2} \right) 
\nn\\
&\
+ C_A \Bigg[
\frac{2 \left(67 \nu ^7+804 \nu ^6+3634 \nu ^5+7380 \nu ^4+4723 \nu ^3-5520 \nu ^2-8712
   \nu -2376\right)}{9 \nu  (\nu +1)^2 (\nu +2)^2 (\nu +3)^2}-4 (\Psi(\nu) + \gamma_E)^2
\nn\\
&\
-\frac{8
   \left(2 \nu ^2+9 \nu +11\right) (\Psi(\nu) + \gamma_E)}{(\nu +1) (\nu +2) (\nu +3)}+12
   \Psi'(\nu)-2 \pi ^2
\Bigg]
\nn\\
&\
+ n_f \Bigg[
\frac{-23 \nu ^7-276 \nu ^6-1190 \nu ^5-2376 \nu ^4-1703 \nu ^3+1644 \nu ^2+3060 \nu
   +864}{9 \nu  (\nu +1)^2 (\nu +2)^2 (\nu +3)^2}+\frac{4 \left(\nu ^2+3 \nu +4\right)
   (\Psi(\nu) + \gamma_E)}{(\nu +1) (\nu +2) (\nu +3)}
\Bigg]
\nn\\
&\ + \cO(\e) \,.
  \end{align}
\end{widetext}
Here $\Psi(z)$ is the digamma function $\Psi(z) = \Gamma'(z)/\Gamma(z)$, which is a meromorphic function with poles at non-positive integer values. For the first few positive integer values of $\nu$, we find
\begin{align}
  \label{eq:jet_one_loop_explicit}
   2 J_1^{q,[1]} = &\
 2 J_1^{g,[1]} =  0
 \,,
\nn\\
 2^2 J_1^{q,[2]} = &\ C_F \left(-\frac{3}{\epsilon }  - \frac{37}{3}  \right) \,,
\nn\\
 2^2 J_1^{g,[2]} = &\ -\frac{14 C_A}{5 \epsilon }-\frac{n_f}{5 \epsilon }
-\frac{898 C_A}{75}-\frac{14 n_f}{25}
 \,,
\nn\\
 2^3 J_1^{q,[3]} = &\ C_F \left(-\frac{9}{2 \epsilon } -\frac{37}{2}  \right) \,,
\nn\\
 2^3 J_1^{g,[3]} = &\ -\frac{21 C_A}{5 \epsilon }-\frac{3 n_f}{10 \epsilon }
-\frac{449 C_A}{25}-\frac{21 n_f}{25}
\,,
\nn\\
 2^4 J_1^{q,[4]} = &\ C_F \left(-\frac{83}{15 \epsilon } -\frac{5206}{225}  \right) \,,
\nn\\
 2^4 J_1^{g,[4]} = &\ 
-\frac{181 C_A}{35 \epsilon }-\frac{38 n_f}{105 \epsilon }
-\frac{82589 C_A}{3675}-\frac{11317 n_f}{11025}
\,,\nn
\end{align}
where we have set $\mu = \sqrt{x_L} Q$ for simplicity. The $\nu = 2$ results agree with Ref.~\cite{Dixon:2019uzg} up to the $2^{-\nu}$ normalization. 

In addition to QCD, we give results for the jet function in $\cN = 4$ SYM. In this theory the one-loop jet function is obtained from a single universal splitting kernel, $P_{\cN = 4} = 2 N_c/(z (1-z))$, and turns out to be quite simple
\begin{gather}
  2^\nu J_1^{\cN = 4, [\nu]} = - 8 N_c (\Psi(\nu) + \gamma_E) \left( \frac{1}{\e} - \ln \frac{x_L Q^2}{\mu^2} \right)
\nn\\
-4 N_c [ \pi^2 + 2 (\Psi(\nu) + \gamma_E)^2 - 6 \Psi'(\nu) ] + \Ord(\e) \,.
  \label{eq:jet_Neq4}
\end{gather}
In both QCD and $\cN = 4$ SYM the jet functions cross zero at $\nu = 1$ due to the conservation of the energy-momentum tensor. This can also be understood from the momentum conservation sum rule in final-state fragmentation.  The constants in the $\cN=4$ jet function exhibit a uniform transcendental weight,\footnote{We assign a transcendental weight $n+1$ to $\Psi^{(n)}$. We also assign transcendental weight $1$ to $\gamma_E$, $\pi$, and $1/\e$. } and by comparing the result in $\cN=4$ with the result for the gluon jet function in QCD, we see that the principal of maximal transcendentality holds (This was already observed for $\nu =2$~($j=3$) in \cite{Dixon:2019uzg}). In the collinear limit, the jet function is determined by a fixed value of the spin, and the harmonic sums (polygamma functions once analytically continued) evaluate to rational numbers obscuring the weight information (see \eqn{eq:jet_one_loop_explicit}). By viewing the observable as a function of $\nu$, we are able to manifest the uniform transcendentality in the collinear limit. We conjecture that uniform transcendentality persists to all orders in $\alpha_s$, since it is ultimately inherited from the uniform transcendental weight of the universal structure constants in \cite{Eden:2012rr}. Uniform transcendentality has also been observed for the DIS structure functions in \cite{Bianchi:2013sta}.

\subsection{LL Resummation and Interpretation}
\label{sec:interp}

In this section, we perform the LL resummation of the $\nu$-correlators in the small angle limit, which provides some intuition for the behavior of the projected correlators as a function of $\nu$. Since the factorization formula and renormalization group evolution equations are straightforward generalizations of those presented in \cite{Dixon:2019uzg}, it is trivial to solve them in an identical manner for generic values of $\nu$, and so we do not discuss this aspect further. We will consider both the case of the conformal $\cN=4$ SYM theory, as well as QCD where there is a non-vanishing $\beta$ function. Resummation at NLL and numeric results will be presented in \Sec{sec:NLL}.

We begin by considering the case of $\cN=4$ SYM. We find that the resummed result for the cumulant $\Sigma^\supnu$ is given by
\begin{equation}
\Sigma^{[\nu]}(x_L) = C^{[\nu]}(\alpha_s) \, x_L^{\gamma_{J^{[\nu]}}^{{\cal N}=4}(\alpha_s)} \,,
\label{Neq4NNLLcumulant}
\end{equation}
where $C^\supnu$ is the structure constant, and
\begin{align}
 \gamma_{J^{[\nu]}}^{{\cal N}=4}(\alpha_s) = \gamma_S^{{\cal N}=4}(\nu+1,\alpha_s)\,,
 \end{align}
and $\gamma_S^{{\cal N}=4}(\nu+1,\alpha_s)$ is the universal local twist-$2$ spin-$\nu+1$ anomalous dimension in $\cN=4$ (To maintain continuity between QCD and $\cN=4$, we use conventions for the anomalous dimensions in $\cN=4$ where $\gamma_{\text{uni}}^{(0)}(j) \propto S_1(j-2)$. It is also common in $\cN=4$ to shift $j$ by two units.). The power-law behavior is due to conformal symmetry~\cite{Hofman:2008ar,Kologlu:2019mfz}, or reciprocity~\cite{Korchemsky:2019nzm,Dixon:2019uzg}. 
Differentiating in $x_L$, we have
\begin{equation}
\frac{d\sigma^{[\nu]}}{dx_L} =  C^{[\nu]}(\alpha_s) \gamma_{J^{[\nu]}}^{{\cal N}=4}(\alpha_s) \, \frac{x_L^{\gamma_{J^{[\nu]}}^{{\cal N}=4}(\alpha_s)}}{x_L} \,,
\label{Neq4differential}
\end{equation}
The scaling of the $\nu$-correlator therefore allows one to probe the spectrum of the underlying field theory through the scaling in the $x_L$ variable. 

The fact that we are able to have a scaling observable for all values of $\nu$ allows us to connect different physical regions with the same observable. In particular, the $\nu$-correlators have different behavior depending on whether $\nu>1$, or $\nu<1$.  For positive integers the scaling anomalous dimensions correspond to the anomalous dimensions of local twist-2 operators, which in a unitary CFT are guaranteed to be positive \cite{Ferrara:1974pt,Mack:1975je}. This implies that the resummed cumulant vanishes as $x_L\to 0$. For $\nu>1$ non-integer values, there is no longer a correspondence with local operators, but the anomalous dimensions remain positive by continuity and monotonicity. For $\nu=1$ (which corresponds to $j=2$), the scaling anomalous dimension vanishes to all orders in perturbation theory, since it corresponds to the anomalous dimension of the stress tensor. This has the interesting consequence that the cumulant is independent of the scaling variable (i.e. the distribution is a $\delta$-function).  For $\nu<1$, the scaling anomalous dimension is negative. In this region there is no correspondence with a local operator, and therefore the standard unitarity bounds do not apply. In particular, this implies that the cumulant diverges as $x_L \to 0$. While this is perhaps unusual for jet observables, this behavior is physical. Some intuition can be gained by recalling the expression for the multiplicity in an $e^+e^-$ collision at a scale $Q^2$
\begin{align}
n(Q^2, \Lambda)\propto \left(  \frac{Q}{\Lambda} \right)^{-2\gamma(\nu=0)}\,,
\end{align}
where $\Lambda$ is an infrared resolution (Recall that in our conventions, $\gamma(1)<0$, which is opposite to the conventions often used when discussing multiplicity). This has a similar behavior to the cumulant in \eqn{Neq4NNLLcumulant}, if we associate $x_L$ with an infrared regulator, leading to an interpretation of the $\nu$-correlator in the $\nu\to 0 $ limit.  Multiplicity correlators have been considered in \cite{Ochs:1988ky,Ochs:1992gd,Ochs:1994rt} (for a recent measurement see \cite{Badea:2019vey}) and the divergence at small angles is well known. We therefore find that the $\nu$-correlators are able to connect, in a single analytic observable, the EEC and multiplicity, as well as all other observables lying in the complex $\nu$-plane.

Moving beyond a conformal theory, at LL accuracy the simple scaling behavior is only modified through the inclusion of the running coupling, and therefore much of the intuition from the case of a conformal theory carries over. In addition to the beta function, in QCD one must also incorporate non-trivial flavor mixing. At LL, the solution of the jet function evolution to the hard scale in QCD is
\begin{align}
  \label{eq:jetLLresum}
  \vec{J}_{\rm LL}^{\,[\nu]} = 2^{-\nu}(1,1) 
\exp \left(- \frac{\widehat{\gamma}^{(0)} (\nu+1)}{\beta_0} \ln \frac{\alpha_s(\sqrt{x_L} Q)}{\alpha_s(\mu)} \right) \,,
\end{align}
where $\widehat{\gamma}^{(0)}(j)$ is the Mellin moment of the singlet timelike
splitting function at LO in QCD, 
$\widehat{\gamma}(j,\alpha_s) = -  \int_0^1 dz\, z^{j-1} \widehat{P}(z,\alpha_s)$, 
where $\widehat{P}(z,\alpha_s)$ is the regularized singlet timelike splitting kernel. Explicitly, 
\begin{align}
  \label{eq:qcdgamma}
  \widehat{\gamma}(j) = 
\begin{pmatrix}
\gamma_{qq}^{(0)}(j) & 2 n_f \gamma_{ qg}^{(0)}(j)
\\
\gamma_{gq}^{(0)}(j) & \gamma_{gg}^{(0)}(j)
\end{pmatrix} \,,
\end{align}
where~\cite{Ellis:1991qj}
\begin{align}
  \label{eq:QCDAD}
  \gamma_{ qq}^{(0)}(j)&\ = -2 C_F \left[ \frac{3}{2} + \frac{1}{j (j+1)} - 2 (\Psi(j+1) + \gamma_E ) \right] \,,
\nn\\
\gamma_{gq}^{(0)}(j)&\ = -2 C_F \frac{ (2 + j + j^2)}{j (j^2 - 1)} \,,
\nn\\
\gamma_{gg}^{(0)}(j)&\ = -4 C_A \bigg[ \frac{1}{j (j-1)} + \frac{1}{(j+1) (j+2)}
\nn\\
&\ \quad  - (\Psi(j+1) + \gamma_E)  \bigg] 
- \beta_0 \,,
\nn\\
\gamma_{qg}^{(0)}(j) &\ = - \frac{(2 + j + j^2)}{j (j+1) (j+2)} \,,
\end{align}
and $\beta_0 = 11/3 C_A - 2/3 n_f$. One can check that the pole terms on the RHS of Eq.~\eqref{eq:jet_cst_one_loop} are given by
\begin{equation}
  \label{eq:pole1}
 - \gamma_{qq}^{(0)} - \gamma_{gq}^{(0)} \quad \text{and} \quad  - \gamma_{gg}^{(0)} - 2 n_f \gamma_{qg}^{(0)} \,,
\end{equation}
respectively~(recall $j = \nu + 1$). 
We therefore see that even in QCD the scaling is still driven by the twist-2 spin-$\nu+1$ anomalous dimensions, however, this behavior is no longer a power law due to the running coupling. This jet function must then be projected on to an appropriate tree level hard function. For example, for the case of $e^+e^-$ one has
\begin{align}
  \label{eq:HLL}
  \vec{H}_{\rm LL} (x) = 
 2  \begin{pmatrix}
    \delta(1-x)
\\
    0
  \end{pmatrix} \,.
\end{align}

\subsection{NLL Resummation and Numerical Results}
\label{sec:NLL}

\begin{figure*}[ht!]
\centering
\subfloat[]{\label{fig:smallXe}
\includegraphics[width=0.3\textwidth]{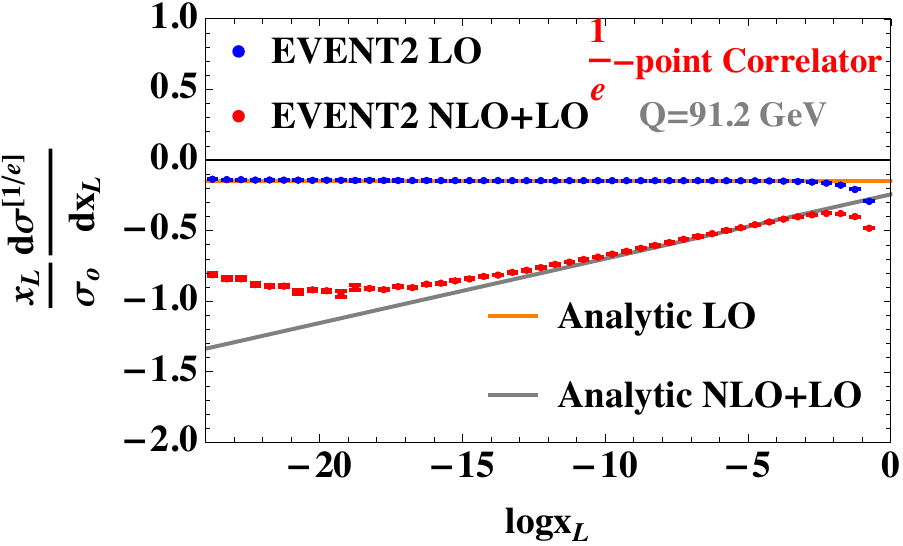}
}
\subfloat[]{\label{fig:smallXpi}
\includegraphics[width=0.3\textwidth]{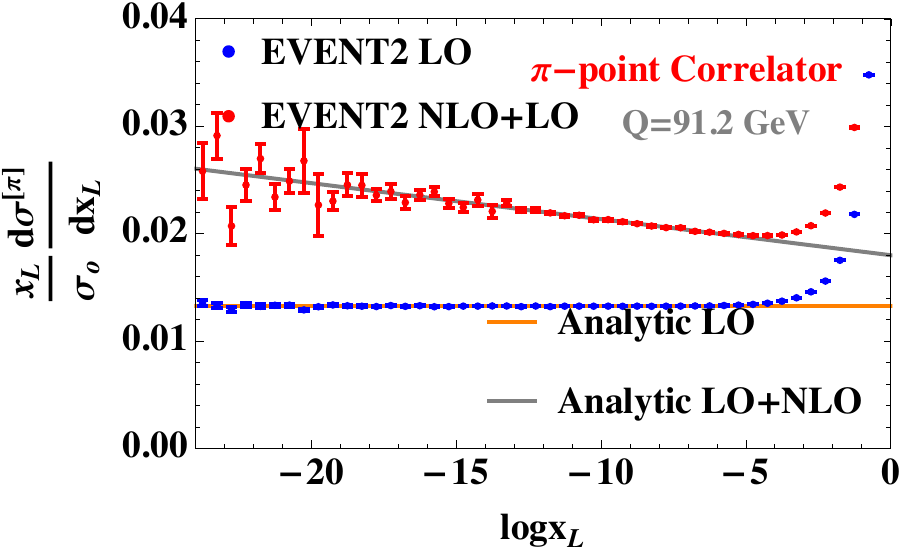}
}
\subfloat[]{\label{fig:smallXcomplex}
\includegraphics[width=0.3\textwidth]{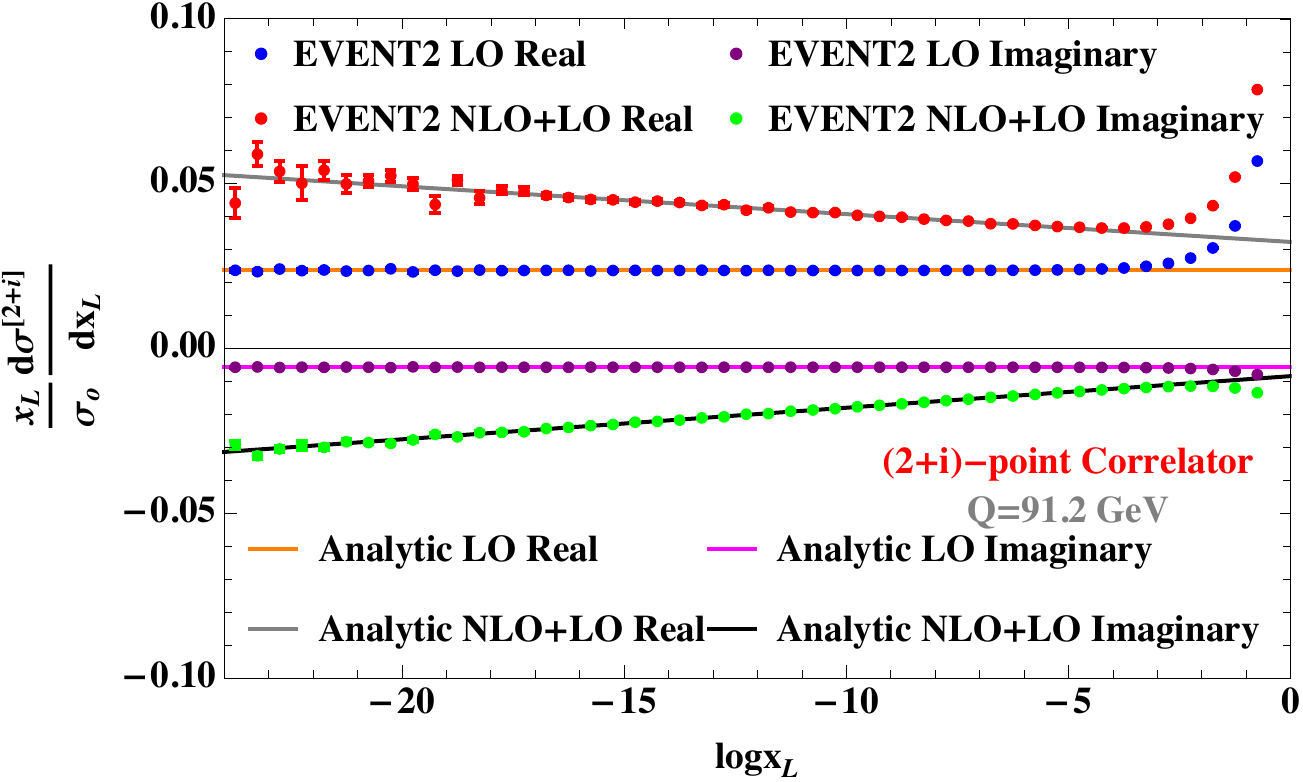}
}
\caption{Comparison between our factorization formula and \texttt{Event2} in the asymptotically small $x_L$ region for $\nu= 1/e \,, \pi \,, 2+i$.}\label{fig:asymptotics}
\end{figure*}

In this section we present several numerical results to highlight interesting features of the $\nu$-correlators, as well as to verify our factorization formula against numerical fixed-order calculations. We leave more detailed phenomenological studies to a future publication.

Results in this section are presented to NLL accuracy,\footnote{Throughout this section we use the logarithmic counting appropriate for single logarithmic observables. Often in jet substructure, a logarithmic counting appropriate for double logarithmic observables is used, even if the observable is single logarithmic. In particular, our NLL result has the same logarithmic accuracy as the NNLL result for the groomed jet mass \cite{Frye:2016aiz,Frye:2016okc}. For the case $N=2$, NNLL resummation is also available \cite{Dixon:2019uzg}, and the relevant anomalous dimensions for the groomed jet mass have also recently been extracted to enable an NNLL prediction \cite{Kardos:2020ppl}.} which resums terms through to $\alpha_s^n \ln(x_L)^{n-1}$. To achieve this accuracy, we need the two-loop timelike splitting functions and QCD beta function, as well as the one-loop hard and jet functions. Only the jet function is new and was given in 
Eq.~\eqref{eq:jet_cst_one_loop}. With these ingredients, numerical predictions in the collinear limit can be obtained using the factorization formula in Eq.~\eqref{eq:fac_nu}, combined with the renormalization group equations in Eqs.~\eqref{eq:hard_evo}, \eqref{eq:jet_evo}.

\begin{figure}[ht!]
  \centering
  \includegraphics[width=0.4\textwidth]{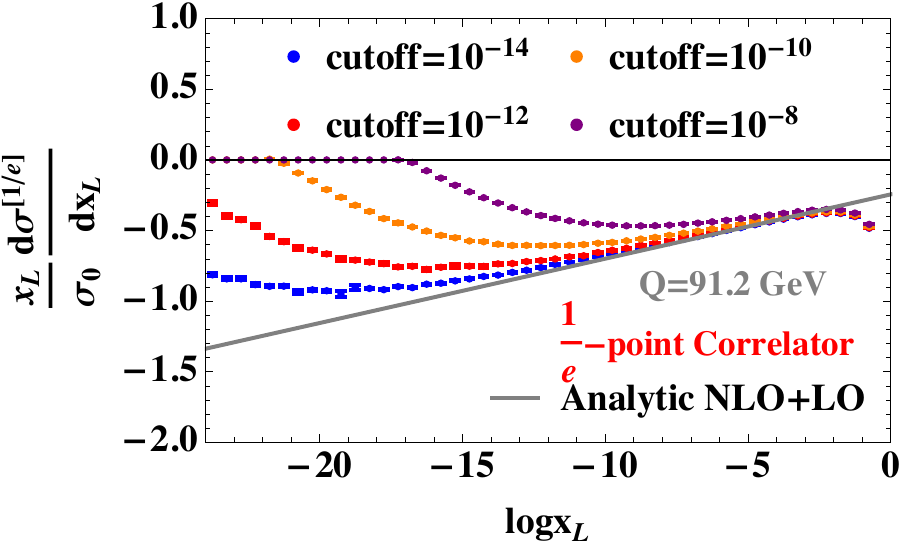}
  \caption{$1/e$-point correlator computed at the NLO with \texttt{Event2} and with different values of \texttt{cutoff} variable. The agreement between \texttt{Event2} and our analytic prediction is better for smaller \texttt{cutoff}.}
  \label{fig:cutoff}
\end{figure}

We first verify the factorization formula in \eqref{eq:fac_nu} by comparing our predictions, truncated to $\Ord(\alpha_s^2)$, with a numerical fixed-order calculation in the small angle limit. We give the results obtained from expanding our factorization formula for three representative values of $\nu$, $\nu = 1/e\,, \pi \,, 2 + i$, in \eqref{eq:fixed_order},
\begin{align}
  \label{eq:fixed_order}
 &\ \frac{x_L}{\sigma_0} \frac{d \sigma^{[e^{-1}]}}{dx_L} = 
a_s (-15.6168)  + a_s^2 \left(516.646 \ln x_L-1107.8\right)  \,,
\nn\\
&\  \frac{x_L}{\sigma_0} \frac{d \sigma^{[\pi]}}{dx_L} = 
a_s 1.41007 + a_s^2 \left(53.8777\, -3.8045 \ln x_L \right) \,,
\nn\\
&\  \frac{x_L}{\sigma_0} \frac{d \sigma^{[2+i]}}{dx_L} = 
+(2.51547\, -0.610332 i) a_s
\\
&\hspace{-0.2cm}
+ a_s^2 \left((98.1462\, -30.7351 i)-(9.58163\, -10.8713 i) \ln
   \left(x_L\right)\right) \,, \nn
\end{align}
where $a_s = \alpha_s/(4 \pi)$. These results are shown in Fig.~\ref{fig:asymptotics} as solid lines. We have also computed the $\nu$-correlator with the QCD event generator \texttt{Event2}~\cite{Catani:1996jh,Catani:1996vz} using the definition given in Eq.~\eqref{eq:projected_mom_nu}. \texttt{Event2} calculates not just the leading power terms in the $x_L\to 0$ limit that are described by the factorization formulas presented in this paper, but also the power suppressed contributions. At small $x_L$, it is expected that the leading power logarithmic terms dominate. The \texttt{Event2} results are shown in Fig.~\ref{fig:asymptotics} as dotted lines. It can be seen that there is agreement between our factorization prediction and \texttt{Event2} when $x_L$ is sufficiently small so that power suppressed terms can indeed be neglected. There is some deviation at NLO for $\nu = 1/e$ when $x_L < e^{-13}$. We believe that this is due to the nature of $\nu < 1$ such that the observable is increasingly sensitive to soft physics, while in \texttt{Event2} there is an un-physical IR cutoff to ensure numerical stability.
To understand the origin of the instability better, we have computed the $\nu=1/e$ correlator in \texttt{Event2} for a number of different internal \texttt{cutoff} variable ranging from $10^{-8}$ to $10^{-14}$. The results show a strong dependence on the cutoff variable, as can be seen clearly from Fig.~\ref{fig:cutoff}. This confirms our belief that the deviation of \texttt{Event2} from our analytic prediction is indeed due to the un-physical IR cutoff in \texttt{Event2}.
\begin{figure*}[ht!]
\centering
\subfloat[]{\label{fig:EinveC}
\includegraphics[width=0.45\textwidth]{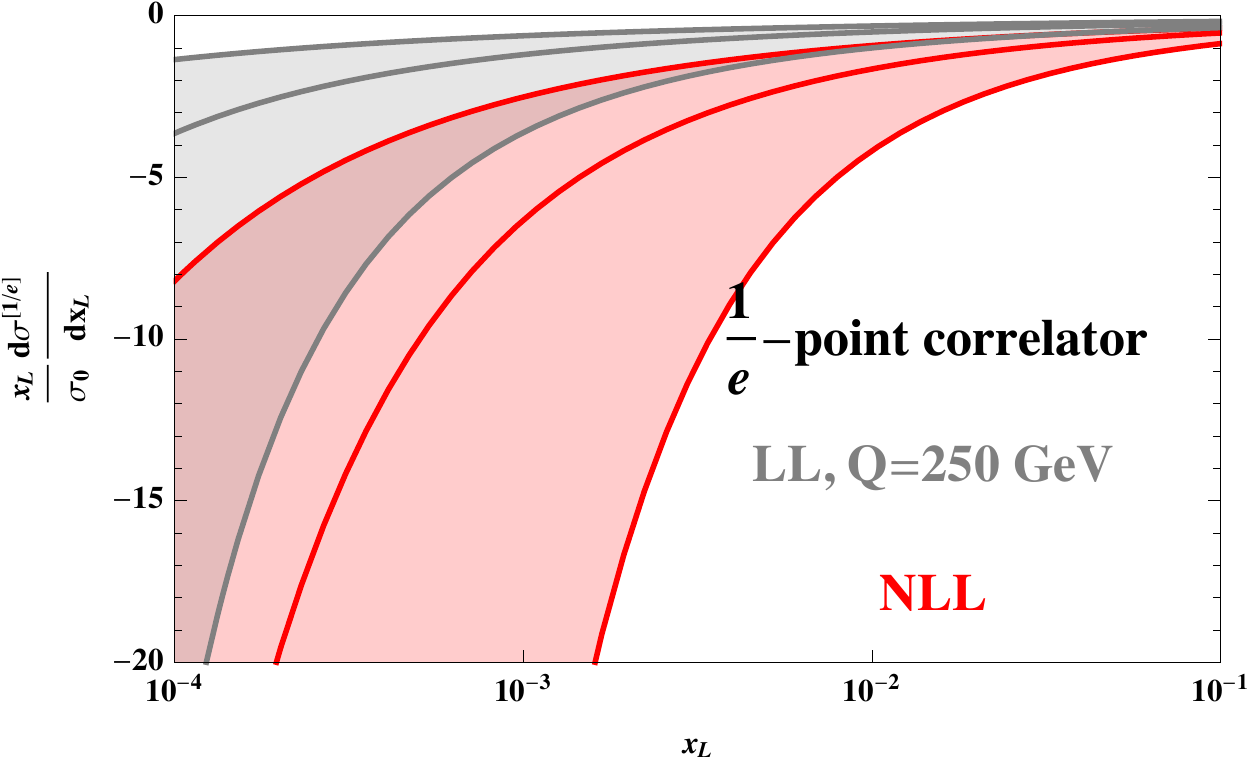}
}
\subfloat[]{\label{fig:EpiC}
\includegraphics[width=0.45\textwidth]{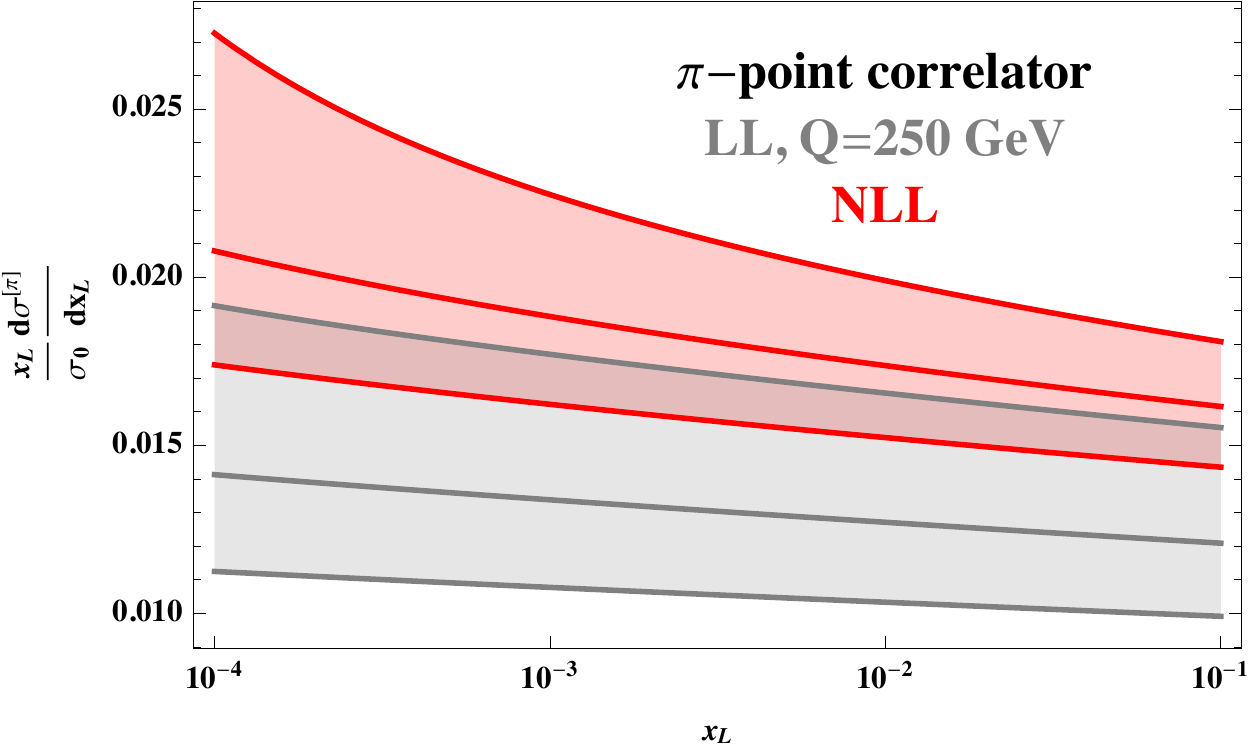}
}
\\
\subfloat[]{\label{fig:E2iCreal}
\includegraphics[width=0.45\textwidth]{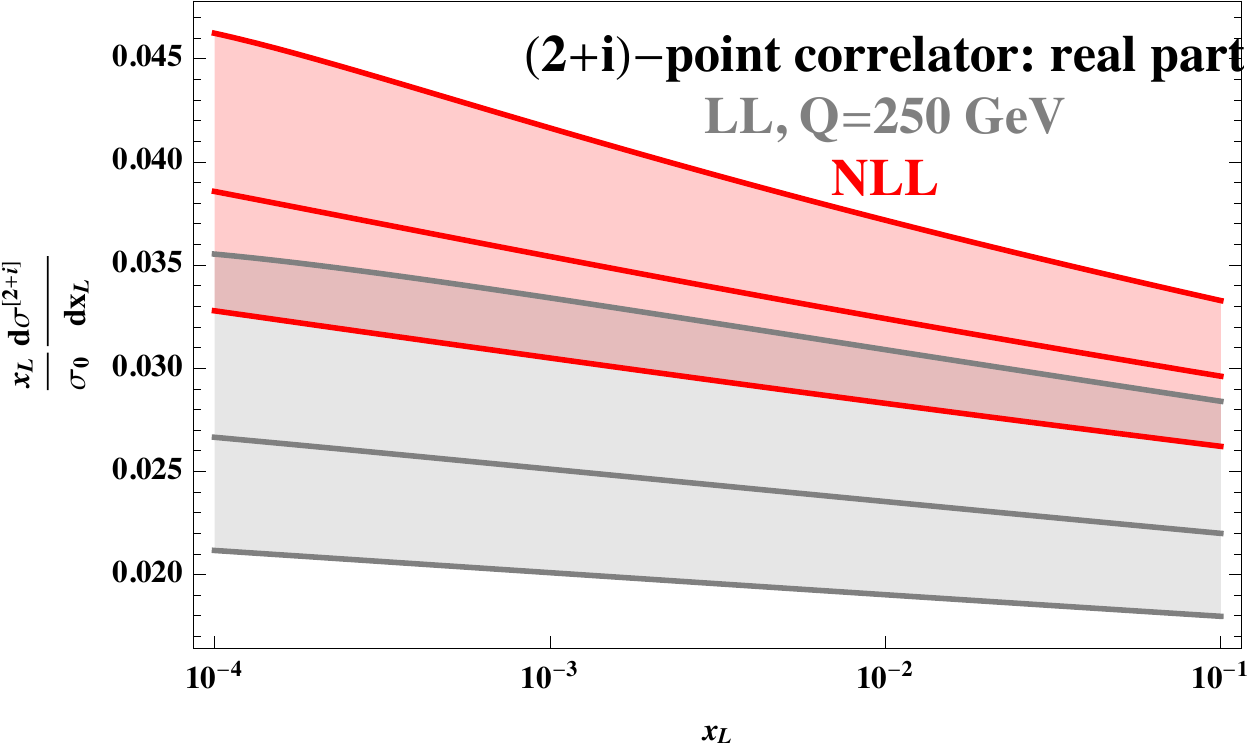}
}
\subfloat[]{\label{fig:E2iCim}
\includegraphics[width=0.45\textwidth]{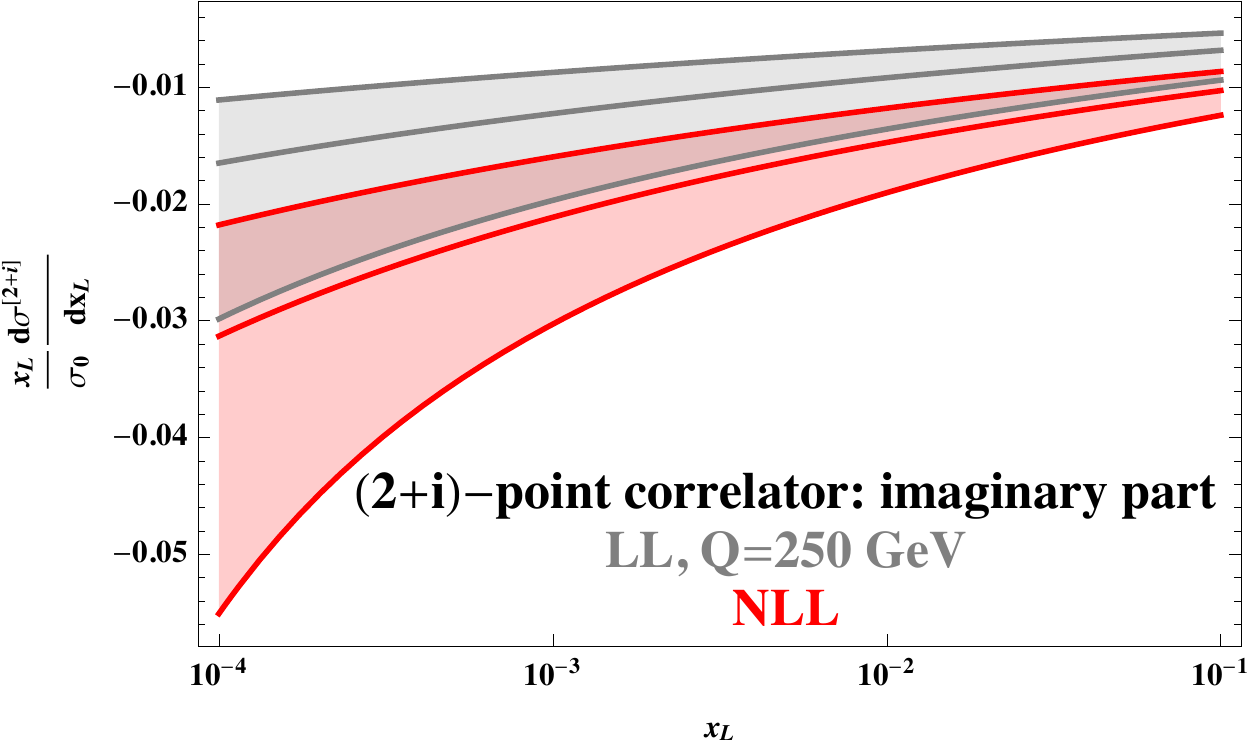}
}
\caption{LL~(grey) and NLL~(red) resummation for $\nu = e^{-1}$, $\pi$, and $2+i$. Scale uncertainties are estimated by varying $\mu$ around $Q$ by a factor of $5^{-1}$ and $5$.}\label{fig:NLL}
\end{figure*}

For $\nu = 2 + i$ the results contain both real and imaginary parts. This is not a problem since what we computed are correlation functions, or weighted cross sections, and do not correspond to probabilities. The agreement between Eq.~\eqref{eq:fixed_order} and \texttt{Event2} provides a strong check on our factorization formula \eqref{eq:fac_nu}. Furthermore, this agreement between the NLO result in \texttt{Event2} and our analytic result, strongly suggests that this observable is IRC safe at $\Ord(\alpha_s^2)$, at least for $0 < x_L < 1$. It would be interesting to have a (dis-)proof of IRC safety for the $\nu$-correlator to all orders. 

We emphasize the different behavior as $x_L\to 0$ for $\nu=e^{-1}<1$, and $\nu=\pi>1$. As discussed above, for $\nu>1$, the behavior (at least in the conformal case) is driven by an anomalous dimension of a local operator which is constrained to be positive in a unitary theory. For $\nu<1$, this association is lost, and the scaling flips sign. This is clearly seen in the behavior of the fixed order calculations in Fig.~\ref{fig:asymptotics}. It also persists once resummation is included.

To resum the large logarithms arising in the collinear limit, we follow the approach of \cite{Dixon:2019uzg} by solving the RG equation Eq.~\ref{eq:jet_evo} iteratively to high orders. We keep the first $50$ terms in $\alpha_s$ expansion, which is sufficient to have convergence to better than one per mille for the range of $x_L$ considered here. In Fig.~\ref{fig:NLL} we depicted the $\nu$-correlator at LL and NLL for $\nu = e^{-1}$, $\pi$, and $2 + i$. These values allow us to emphasize the qualitatively different behavior for $\nu>1$ and $\nu<1$. We set $Q = 250$ GeV so that we have sufficiently large window at small $x_L$ for perturbative evolution. We observe reasonable perturbative convergence when going from LL to NLL for $\nu = \pi$ and $2 + i$. On the hand, the convergence is bad for $\nu = e^{-1}$, as indicated by the non-overlapping scale bands. Since for $\nu < 1$, the $\nu$-correlator probes the small $x$ fragmentation kernel, one might expect that some form of small $x$ resummation for the anomalous dimension becomes necessary,  which we leave for future work. The scale uncertainties are still large at NLL due to large perturbative corrections to the NLL coefficients, as already observed in \cite{Dixon:2019uzg} for EEC. This calls for an NNLL calculation (which was already performed in \cite{Dixon:2019uzg} for the case of $N=2$) for generic values of $\nu$, which we leave for future work. 

\begin{figure}[ht!]
  \centering
  \includegraphics[width=0.45\textwidth]{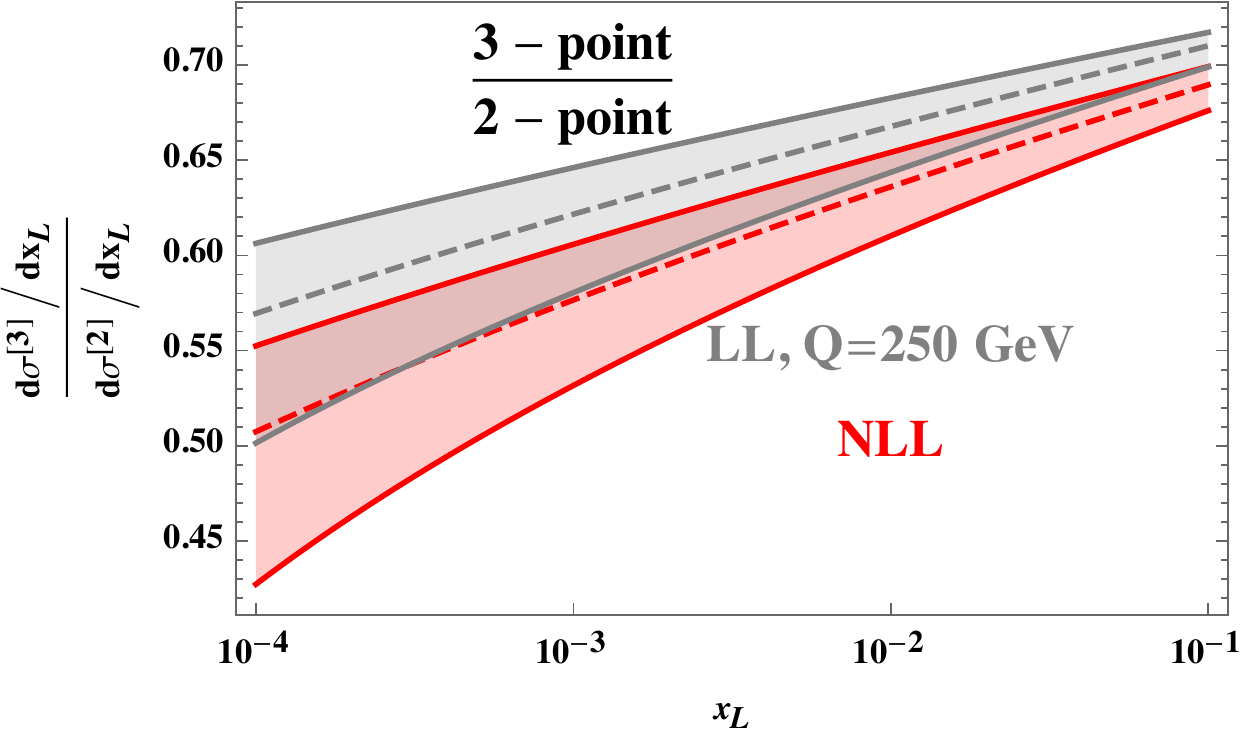}
  \caption{Ratio of $3$-point and $2$-point correlator at LL and NLL. Scale uncertainties are estimated by varying $\mu$ in the numerator and denominator simultaneously by of factor of $5$ and $5^{-1}$.}
  \label{fig:ratio}
\end{figure}

\begin{figure}[h!]
  \centering
  \includegraphics[width=0.45\textwidth]{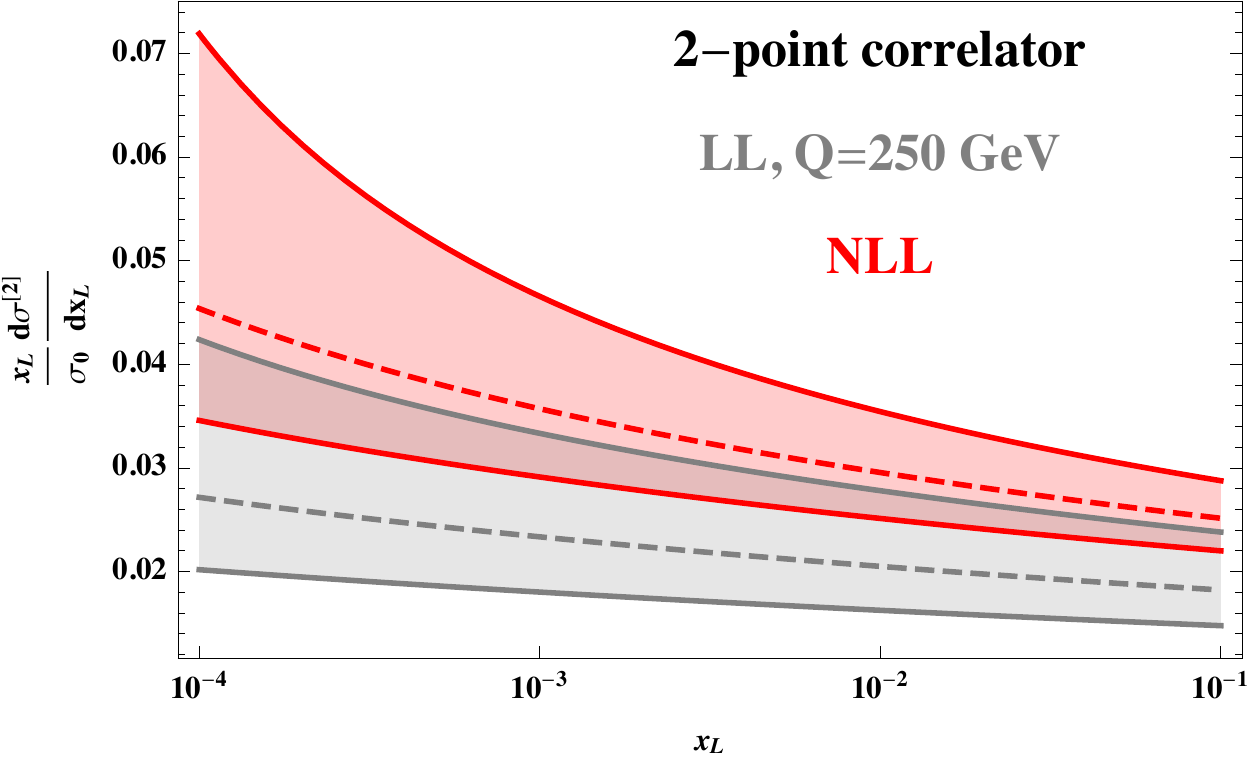}
\\
  \includegraphics[width=0.45\textwidth]{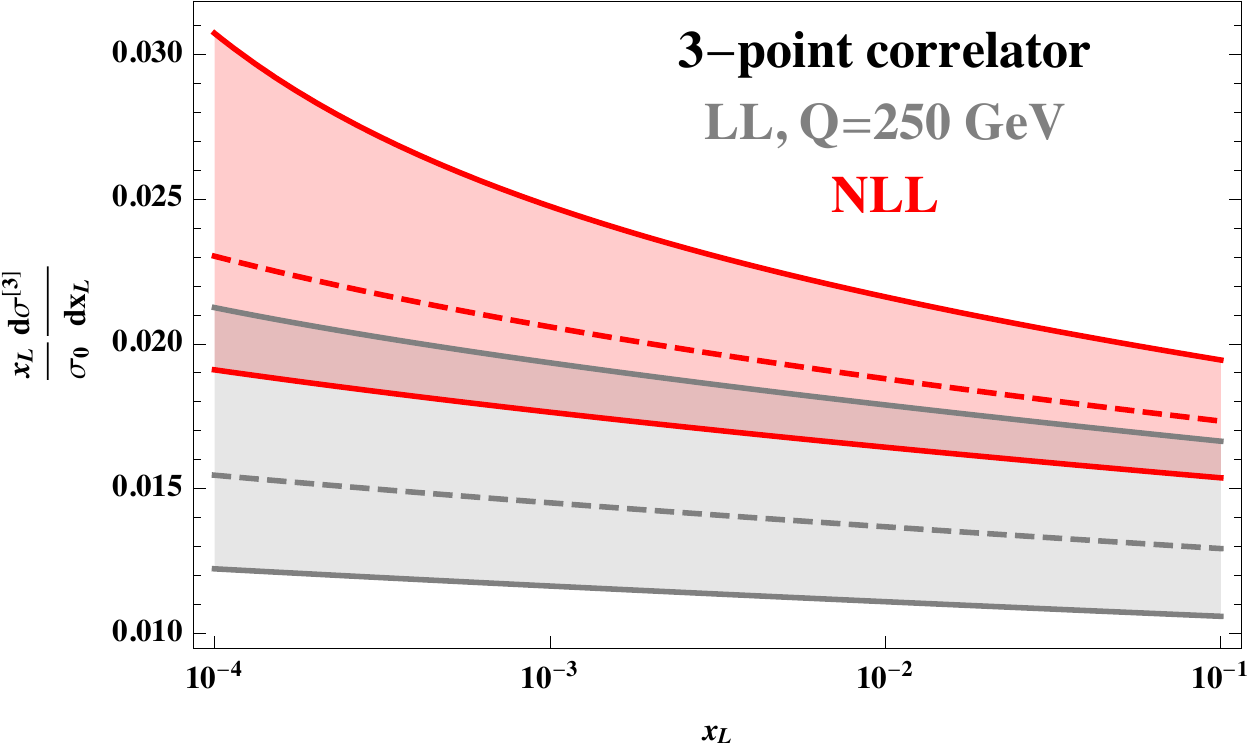}
  \caption{$2$-point and $3$-point correlators at LL and NLL. Scale uncertainties are estimated by varying $\mu$ in the numerator and denominator simultaneously by of factor of $5$ and $5^{-1}$.}
  \label{fig:two_three}
\end{figure}

In Fig.~\ref{fig:ratio} we plot the ratio observable, $d\sigma^{[3,2]}/dx_L$ at LL and NLL. As a comparison, we also plot the $2$-point correlator~(EEC) and $3$-point correlator in Fig.~\ref{fig:two_three}. As is advocated in subsection \ref{sec:observables_b}, it can indeed be seen that the ratio observable dramatically reduce the scale uncertainties and the magnitude of the corrections when going from LL to NLL. While the remaining scale uncertainty is still large, it nevertheless gives us hope that by going to NNLL one would be able to control the perturbative uncertainties.

\section{Resummation for Track Correlators}
\label{sec:N_track}

In this section we briefly describe the resummation of the energy correlators measured on tracks.  The goal of this section is to illustrate that track functions interface naturally with energy correlators, since much like the energy correlators, moments of the track functions evolve with moments of the twist-2 spin-$j$ anomalous dimensions.  In this sense we view EECs as the natural observable for tracks. In this section we will consider the specific case of the two point correlator at leading logarithmic accuracy. 

As with the case of the energy correlators we can write down a timelike factorization formula for the cumulant, $ \Sigma_{\text{tr}}^\supnu$ of $\nu$-point correlators measured on tracks (here the subscript tr denotes tracks)
\begin{align}
  \label{eq:fac_nu_tracks}
  \Sigma_{\text{tr}}^\supnu(x_L, \ln\frac{Q^2}{\mu^2}) = \int_0^1 \! dx\, 
x^\nu \vec{J}_\text{tr}\,^\supnu(\ln \frac{x_L x^2 Q^2}{\mu^2}) \cdot \vec{H}(x, \frac{Q^2}{\mu^2}) \,.
\end{align}
Crucially, the incorporation of tracks does not change the hard function, and only enters into the jet function. Renormalization group consistency then fixes that the evolution of the jet function on tracks is identical to the evolution of the standard jet function
\begin{align}
  \label{eq:jet_evo_track}
  \frac{d \vec{J}_\text{tr}\,^\supnu( \ln \frac{x_L Q^2}{\mu^2})}{d \ln\mu^2}
=
\int_0^1 \! dy\, y^\nu \vec{J}_\text{tr}\,^\supnu( \ln \frac{x_L y^2 Q^2}{\mu^2})
\cdot 
\widehat{P}(y) \,.
\end{align}
This constraint arises from the fact that the energy correlators are collinear (single logarithmic) observables so that the factorization formula consists of only two functions.  This leads to a significant simplification as compared to the case of Sudakov (soft sensitive) observables that exhibit a factorization into hard function that is independent of the measurement, and two functions, the jet and soft functions that depend on the measurement. In the case of a typical Sudakov observable, the use of tracks modifies both the jet and soft functions in an equal and opposite manner that is not constrained by renormalization group consistency with the hard function. This is the case for the example of track thrust considered in \cite{Chang:2013iba}, where the anomalous dimensions for the jet and soft functions are modified by a non-perturbative constant. For the energy correlators, the anomalous dimensions remain perturbative, and equal to their value without tracks.

We can therefore immediately derive the leading logarithmic result for the $\nu$-correlators measured on tracks
\begin{align}
  \label{eq:jetLLresum_track}
\Sigma_{\text{tr}}^\supnu(x_L)&= 2^{- \nu + 1} (T_q^{(\nu)}(\sqrt{z} Q),T_g^{(\nu)}(\sqrt{z} Q)) \\
& \cdot
\exp\left[- \frac{\widehat{\gamma}^{(0)} (\nu + 1) }{\beta_0} \ln \frac{\alpha_s(\sqrt{x_L} Q)}{\alpha_s(Q)} \right]
   \cdot     \begin{pmatrix}
     1
\\
    0
  \end{pmatrix}\,. \nn
\end{align}
We find the simplicity of this result to be quite remarkable, and suggestive that it can be extended to higher perturbative orders.

Since the goal of this section is to illustrate the interplay between track functions and energy correlators, rather than perform a detailed phenomenological study, we have made the simplification in \eqn{eq:jetLLresum_track} of assuming only one flavor of quark. While the evolution of the track functions for different quarks flavors is the same, the non-perturbative functions are in general distinct (although in reality, they are quite similar, see \cite{Chang:2013iba,Chang:2013rca}). Therefore, in reality one must extend $\vec{J}_\text{tr}$ to include all the flavors separately. However, for notational simplicity we will not consider this complication here.

In \eqn{eq:jetLLresum_track}, the moments of the track functions are evaluated at the scale $\sqrt{x_L} Q$, and therefore to achieve a particular logarithmic accuracy, one must also know the evolution of the track functions to the corresponding order. While the track functions themselves evolve with complicated non-linear evolution equations that are not currently known at higher orders, the moments of the track functions evolve via linear evolution equations. This will also allow us to explain an interesting feature of \eqn{eq:jetLLresum_track}, namely that, taking for concreteness $\nu=2$, $T_q^{(2)}(\sqrt{x_L} Q)$ and $T_g^{(2)}$ appear in the result, even though these should physically only appear as boundary terms. We will see the resolution of this fact due to the tight interconnection between the RG equations for the track functions, and those for the EEC. In the rest of this section we will focus on the particular case of the two point correlator, however the extension to higher points should be clear.

The renormalization group evolution equation for the track function at lowest order is a non-linear evolution equation \cite{Chang:2013iba,Chang:2013rca}
\begin{align}\label{eq:RG_track_func}
\mu \frac{d}{d\mu}T_i(x,\mu)&=\frac{1}{2}\sum\limits_{j,k} \int dz dx_j dx_k \frac{\alpha_s(\mu)}{\pi} P_{i\to jk}(z) \\
&\cdot T_j(x_j,\mu) T_k(x_k,\mu) \delta[x-zx_j-(1-z)x_k]\,. \nn
\end{align}
Little is known about its higher order structure, but it is expected to become increasingly non-linear. A large simplification occurs when one only has to deal with a finite number of moments of the track functions, as occurs for the energy correlators. Taking moments of \eqn{eq:RG_track_func}, we find
\begin{widetext}
\begin{align}
\frac{d}{d\ln \mu^2} T_g^{(n)} = &\ -  \sum \limits_{k=0}^{n-1}
 \begin{pmatrix}
n
\\
k
\end{pmatrix}
\left[  \sum \limits_{i=0}^k
 \begin{pmatrix}
k
\\
i
\end{pmatrix}
(-1)^i \left( \frac{1}{2} \gamma_{gg}(n-k+i+1)T_g^{(n-k)}T_g^{(k)} +\sum_{m=1}^{n_f} \gamma_{qg} (n-k+i+1) T_{q_m}^{(n-k)}T_{q_m}^{(k)}  \right)
 \right]
\nn
\\
&\
-  (\frac{1}{2} \gamma_{gg} (n+1) T_g^{(n)} 
+\sum_{m=1}^{n_f} \gamma_{qg} (n+1) T_{q_m}^{(n)} ) \,,
\end{align}
and
\begin{align}
\frac{d}{d\ln \mu^2} T_{q_m}^{(n)}= - \sum \limits_{k=1}^n  
 \begin{pmatrix}
n
\\
k
\end{pmatrix}
\sum \limits_{i=0}^{n-k}
 \begin{pmatrix}
n-k
\\
i
\end{pmatrix}
(-1)^i \gamma_{gq}(k+i+1) T_{q_m}^{(n-k)}T_g^{(k)} 
- \gamma_{qq}(n+1) T_{q_m}^{(n)} \,,
\end{align}
where we have considered QCD with $n_f$ light flavors. 

In particular, for the first two moments, which are required for the two point energy correlator, we have the RG equations~(for simplicity we show below QCD with a single quark flavor)
\begin{align}\label{eq:one_point_charge}
\frac{d}{d\ln \mu^2} T_g^\one&=-\gamma_{gg}(2) T_g^\one -2 \gamma_{qg}(2) T_q^\one\,, \nn \\
\frac{d}{d\ln \mu^2} T_q^\one&=-\gamma_{qq}(2) T_q^\one -\gamma_{gq}(2) T_g^\one\,, 
\end{align}
and
\begin{align}
\frac{d}{d\ln \mu^2} T_g^\two&=-\gamma_{gg}(3) T_g^\two -2 \gamma_{qg}(3) T_q^\two-\gamma_{gg}(2) T_g^\one T_g^\one +\gamma_{gg}(3) T_g^\one T_g^\one -2 \gamma_{qg}(2) T_q^\one T_q^\one +2 \gamma_{qg}(3) T_q^\one T_q^\one\,, \nn \\
\frac{d}{d\ln \mu^2} T_q^\two&=-\gamma_{qq}(3) T_q^\two -\gamma_{gq}(3) T_g^\two +2\gamma_{gq}(3) T_q^\one T_g^\one-2 T_q^\one T_g^\one \gamma_{gq}(2)\,.
\end{align}
Much like the evolution equations for two point energy correlator, we see that these evolution equations involve the twist-2 spin-$3$ dimensions, although they also involve the spin-$2$ anomalous dimension in off diagonal entries.
We can write this RG as a matrix evolution equation\footnote{Again, we emphasize that here we consider the case of a single quark flavor. The extension to five flavors is straightforward, albeit notationally cumbersome.}
\begin{align}\label{eq:bigmatrix}
\frac{d}{d\ln \mu^2}
 \begin{pmatrix}
T_g^\two\\T_q^\two\\T_q^\one T_q^\one\\ T_g^\one T_q^\one \\T_g^\one T_g^\one
\end{pmatrix}
=
 \begin{pmatrix}
{\color{red}-\gamma_{gg}(3)} && {\color{red}-2 \gamma_{qg}(3)} && 2 \gamma_{qg}(3)-2 \gamma_{qg}(2) &&0 && \gamma_{gg}(3)-\gamma_{gg}(2)\\
{\color{red}-\gamma_{gq}(3)} && {\color{red}-\gamma_{qq}(3)} &&0 && 2\gamma_{gq}(3) -2 \gamma_{gq}(2) &&0 \\
{\color{blue}0}&&{\color{blue}0}&& {\color{red}-2 \gamma_{qq}(2)} &&{\color{red} -2\gamma_{gq}(2)} &&{\color{red}0} \\
{\color{blue}0}&&{\color{blue}0}&&{\color{red}-2 \gamma_{qg}(2) }&&{\color{red} -\gamma_{gg}(2) -\gamma_{qq}(2)}&&{\color{red}-\gamma_{gq}(2)}\\
{\color{blue}0}&&{\color{blue}0}&&{\color{red}0}&&{\color{red}-4\gamma_{qg}(2)} &&{\color{red}-2 \gamma_{gg}(2)}
 \end{pmatrix}
 \begin{pmatrix}
T_g^\two\\T_q^\two\\T_q^\one T_q^\one\\ T_g^\one T_q^\one \\T_g^\one T_g^\one
\end{pmatrix}\,.
\end{align}
\end{widetext}
While the RG for the full track function will become more and more complicated at each perturbative order, the RG for any fixed moment should close, namely the RG for $T_i^{(n)}$, involves only $T_i^{(m)}$ with $m\leq n$. Furthermore, there are two additional features of this matrix that can be derived by considering its interplay with the resummation for the EEC, and that we therefore believe will hold to all orders: lower moments never mix back into the higher moments which fixes the blue entries of the matrix to be zero, and the mixing of highest moments $T_i^{(n)} \to T_j^{(n)}$ (shown by the entries in red) is identical to that of the energy correlators (Note that this holds for the $T_i^\one T_j^\one$ entries of the matrix since the RG for these product terms is derived from the RG for $T_i^\one \to T_J^\one$ mixing.).

While the first of these conditions is easy to understand, the second arises from the fact that for the EEC one should not require contact terms in the bulk of the distribution.
As a simple example to illustrate this, we can consider the case of pure Yang-Mills, as it avoids the need to diagonalize matrices.\footnote{This example is artificial in that pure Yang-Mills does not have charged particles. However, we can formally consider the mixing problem in this theory without specifying the non-perturbative track functions.} In pure Yang-Mills, we have
\begin{align}
\frac{d}{d\ln \mu^2} \begin{pmatrix}
T_g^\two\\T_g^\one T_g^\one
\end{pmatrix}
=
\begin{pmatrix}
-\gamma(3)&&\gamma(3) \\
0&&0
\end{pmatrix}
\begin{pmatrix}
T_g^\two\\T_g^\one T_g^\one
\end{pmatrix}\,.
\end{align}
The LL resummed result in pure Yang-Mills is
\begin{align}
\Sigma_{\text{tr}}^{[2]}(x_L)= \tfrac{1}{2} T_g^{(2)}(\sqrt{x_L} Q) \left(\frac{\alpha_s(\sqrt{x_L} Q)}{\alpha_s(Q)} \right)^{-\frac{\gamma^{(0)}(3)}{\beta_0}}  \,.
\end{align}
This result is naively surprising, since it depends on $T_g^\two$, which should only be required to describe the contact terms at $x_L=0$. However, the resolution to this is that we should also evolve the track function perturbatively to the common scale $Q$. Using the RG, we find that at LL we can rewrite this as
\begin{align}
\Sigma_{\text{tr}}^{[2]}(x_L)= \tfrac{1}{2} [T_g^{(1)} (Q)]^2
\left(\frac{\alpha_s(\sqrt{x_L} Q)}{\alpha_s(Q)} \right)^{-\frac{\gamma^{(0)}(3)}{\beta_0}} \,,
\end{align}
which corresponds with the physical intuition. We therefore find that it evolves with the identical anomalous dimension at LL regardless of whether or not tracks are used. This relies crucially on the fact that the mixing for the track functions is the same as for the energy correlators. For the case where both quarks and gluons are present, one can easily check that the same mechanism occurs using \eqn{eq:bigmatrix} and that all dependence on $T_q^\two$ and $T_g^\two$ cancels at LL accuracy, Furthermore, one finds specific linear combinations, $c_{ij}T^\one_i T^\one_j$, that evolve with the same leading logarithmic anomalous dimensions. One also finds other combinations of tracks functions that vanish when the first moments of the track functions are flavor independent, such as $(T_g^\one T_g^\one -2 T_g^\one T_q^\one +T_q^\one T_q^\one)$, that can evolve with other anomalous dimensions, but that are numerically irrelevant. It would be interesting to study this in more detail with a proper extraction of the track functions, however, we leave this to future work.

We therefore believe that the understanding of the energy correlators places strong constraints on the understanding of the RG evolution of moments of track functions, and that they interplay naturally. To extend the calculation of the EEC or EEEC to higher perturbative orders will require understanding the evolution of moments of the track function to higher perturbative orders. This has not been explored at all, and it will be interesting to understand its consistency. Using the arguments of this section, we believe that the form of the matrix in \eqn{eq:bigmatrix} will persist at higher orders. Only the entries in black are not fixed. We suspect that at higher orders these entries will not correspond to moments of splitting functions, but we believe that they can be straightforwardly calculated by extracting the IR poles from the calculation of the two loop EEC jet function computed on tracks. We therefore believe that the evolution equations for the low moments of the track functions should be much more tractable than the non-linear evolution equations for the full track functions, and that significant insight into their structure can be gained by studying the energy correlators.

\begin{figure}[h!]
  \centering
  \includegraphics[width=0.45\textwidth]{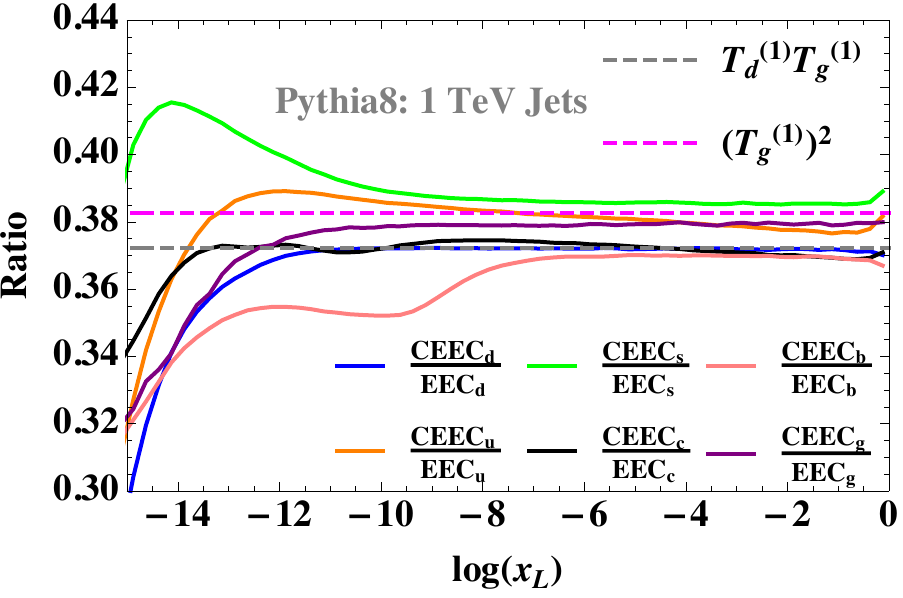}
  \caption{Ratios between two point energy correlators measured on tracks to those using full calorimetric information as computed using Pythia. This illustrates that over a wide perturbative regime the LL evolution is approximately the same with or without tracks, as discussed in more detail in the text.}
  \label{fig:track_verify}
\end{figure}

In \Fig{fig:track_verify} we compare the ratio between the EEC as measured on tracks vs. full calorimetric information for quark and gluon jets, as computed using Pythia.  This is compared with $(T_g^{(1)})^2$ and $T_d^{(1)} T_g^{(1)}$, extracted from \cite{Chang:2013rca}. The flatness of the ratio arises due to the interesting interplay between the anomalous dimensions for the moments of the track functions, and those for the energy correlators. We should emphasize that this comparision should be taken with a grain of salt, since it is sensitive to the precise settings in Pythia (which were presumably not the same in \cite{Chang:2013rca}) as in our study, but is meant to show qualitative agreement. A more detailed analysis, and a calculation at NLL will be presented in future work.

In summary in this section we have emphasized two significant simplifications that arise when studying the resummation of energy correlators measured on tracks. First, the fact that these observables are purely collinear allows the anomalous dimensions of the observables as measured on tracks to be fixed by renormalization group consistency. Secondly, and much more importantly, since the observables involve only a finite number of moments of the track functions, their RG evolution reduces to a linear problem, which is constrained by the structure of the RG for the energy correlators. We believe that these two advantages will enable higher order resummation for track observables, which is a qualitative advance in precision calculations.

A similar story to that presented in this section for tracks also holds for charge correlators. As with the track correlators, one can restrict to the study of their moments, which avoids non-linear evolution equations. It would be interesting to consider these observables in more detail.

\section{Conclusions and Outlook}
\label{sec:conclusions}

In this paper we have advocated for the use of jet substructure observables that are more closely connected to correlation functions of energy (or charge) flow operators in the underlying field theory. In particular, we have shown that there are considerable advantages, both perturbatively and non-perturbatively, to using observables that can be expressed in terms of correlation functions of a finite number of energy flow operators. 

We introduced an infinite family of observables, the projected energy correlators, that project the $N$-point correlators down to a single scaling variable that can be measured experimentally. These observables have simple theoretical properties, allowing for their resummation in the small angle limit for any $N$ at NLL accuracy. This matches the current state of the art resummation accuracy for jet substructure observables, but for an infinite family of observables, and in a single analytic formula. These observables are also amenable to higher order perturbative calculations using modern techniques for loop integrals, which we will consider in future work. Indeed, for the particular case of $N=2$, results at NNLL are already available \cite{Dixon:2019uzg}.

In addition to the perturbative simplicity of the projected energy correlators, we have also shown that observables that can be expressed in terms of a finite number of energy correlators are particularly simple to interface with non-perturbative tracking information. We showed that the $N$-point correlator requires only the knowledge of the $m$-th moments, with $m\leq N$, of the track functions, and that perturbative calculations can be trivially upgraded to calculations on tracks by weighting specific partonic configurations with these moments. This contrasts with calculations of more standard observables on tracks, which involve the complete functional dependence of the track function, and are difficult beyond leading order. The ability to incorporate tracking information is a key advantage of the energy correlators in the LHC environment, but it may also have applications for precision extractions of $\alpha_s$ at $e^+e^-$ colliders. 

A new aspect of our formulation is that it enables an analytic continuation in $N$ of the projected $N$-point correlators. This allows jet observables to explore the complete complex $j$ plane of the twist-2 spin-$j$ operators. For non-integer values of $N$, the $\nu$-correlators correlate infinite combinations of particles within a jet, yet they probe a particularly simple aspect of the underlying physics. This may be a general feature of observables in jet physics, namely that observables with simple physical properties may be algorithmically complex. The analytic continuation of the observables also places them into a clean analytic family of observables that probe particular properties of the collinear limit, and makes manifest certain properties of the result, such as the uniform transcendentality. We believe that this is an important step towards identifying more structure in the physics of jet substructure observables, and it would be interesting to understand other analytic families.

Since the goal of this paper was to introduce the projected energy correlators and highlight some of their convenient theoretical properties, there are a large number of directions for future study. Phenomenologically, an important goal will be to compute the ratio of the three point to two point correlator at NNLL (both with and without track information) at the LHC. All required anomalous dimensions are known (the timelike splitting kernels are known in QCD to NNLO \cite{Mitov:2006wy,Mitov:2006ic,Moch:2007tx,Almasy:2011eq}). In \cite{Dixon:2019uzg} the two-loop jet function for the two point correlator in QCD was obtained using sum rules. To incorporate tracking information, the calculation would need to be done directly, but this should be feasible, as should be the calculation of the NNLO jet function for the projected three point correlator. The primary difficultly at hadron colliders is the hard functions. The hard functions are currently known at NLO \cite{Aversa:1988fv,Aversa:1988mm,Aversa:1988vb,Jager:2002xm} and can be approximated at partonic threshold to higher orders \cite{Hinderer:2018nkb}. Using modern techniques they should be computable to NNLO.

Another aspect of the energy correlators that will be important to understand for phenomenological applications is the structure of their non-perturbative corrections. This should also be considerably simplified for observables defined directly in terms of correlation functions of energy flow operators. The leading non-perturbative corrections for the energy correlators at generic angles were studied in \cite{Korchemsky:1999kt}, where they were found to take a simple form. Furthermore, it has been found that in the small angle limit where they are purely non-perturbative, these observables have simple power law behavior equal to that for an infinitely strongly coupled system \cite{us:open_data}. 

More formally, it will also be interesting to understand in detail the relation of the $\nu$-correlators to the light ray operators of \cite{Kravchuk:2018htv}, which provide the analytic continuation in spin of the standard twist-$2$ spin-$N$ operators. This connection has been explored in detail for the case of $N=2$, but the study of higher integer $N$, as well as non-integer $\nu$ may lead to a better understanding of these jet substructure observables, or facilitate their calculation.

Finally, it would also be interesting to design other observables of this form that are directly related to the underlying energy correlators, for example projections involving two or three variables, instead of the single variable case considered here. This would be a first step towards understanding and organizing the space of jet substructure observables and their relation to the physical operator content of the field theory. Along these lines, observables with energy weighting $E^\kappa$ were considered in \cite{Elder:2017bkd}, giving rise to ``fractal jet observables". It would be interesting to understand more formally what these correspond to in terms of lightray operators, and if these have interesting theoretical properties.

The use of energy correlators for jet substructure opens the door to precision calculations at the LHC, combining high order perturbative calculations with the use of tracking and charge information. It will also facilitate the development of connections between the study of jet substructure and more formal studies of the properties of light ray operators in quantum field theory. We hope to further develop these connections in future work.

\section{Acknowledgements}

We thank Lance Dixon, Ben Nachman, Jesse Thaler, Patrick Komiske, Eric Metodiev, Wouter Waalewijn, David Simmons-Duffin, Petr Kravchuk, Cyuan Han Chang, Phil Harris, Matt Strassler, Gregory Korchemsky, Yixiao Liu, Ming-xing Luo, Tong-Zhi Yang, Duff Neill, and Andrew Larkoski for useful discussions, encouragement and comments on the manuscript, and in particular Wouter Waalewijn for interesting us in the possibility of considering the use of tracks. I.M. thanks the MIT CTP for its hospitality and support while portions of this work were performed. H.C., X.Y.Z and H.X.Z. are supported in part by the National Natural Science Foundation of China under contract No. 11975200. I.M. is supported by the Office of High Energy Physics of the U.S. DOE under Contract No. DE-AC02-76SF00515.

\bibliography{EEC_forward.bib}{}
\bibliographystyle{apsrev4-1}


\newpage

\onecolumngrid
\appendix

\section*{Appendix}

\subsection{Hard Functions}\label{app:hard_func}

In this Appendix we collect the hard functions used in the factorization formula for the projected energy correlators in the collinear limit.

\begin{widetext}
For $e^+e^-$ annihilation, the hard function is given by
\begin{align}
  \label{eq:hard_ee}
\frac{1}{2}  H_q^{\rm ee} = &\ \delta(1 - x) + \frac{\alpha_s}{4 \pi} C_F 
\Bigg[
\left(\frac{4 \pi ^2}{3}-9\right) \delta(1-x)
\nn\\
&\
+4 \left[\frac{\ln(1-x)}{1-x} \right]_++\left(4 \ln
   (x)-\frac{3}{2}\right) (2 \frac{1}{[1-x]}_+-x-1)-\frac{9 x}{2}+2 (-x-1) \ln
   (1-x)+\frac{7}{2}
\Bigg] + \cO(\alpha_s^2) \,,
\nn\\
H_g^{\rm ee} = &\ C_F \Bigg[
\frac{4 \left(x^2-2 x+2\right) \ln (1-x)}{x}+\frac{8 \left(x^2-2 x+2\right) \ln   (x)}{x}
\Bigg] + \cO(\alpha_s^2) \,.
\end{align}
For Higgs decay via the effective $hgg$ operator, the hard function is given by
\begin{align}
  \label{eq:hard_h}
\frac{1}{2}  H_g^{\rm h} = &\ \delta(1-x) + \frac{\alpha_s}{4 \pi} C_A
\Bigg[
\frac{11}{3} \left(-\frac{1}{[1-x]}_++x^2+x+1\right)+(4 \ln (1-x)+8 \ln (x))
   \left(\frac{1}{[1-x]}_+-x^2+x+\frac{1}{x}-2\right)
\nn\\
&\
+\left(\frac{67}{9}+\frac{4 \pi
   ^2}{3}\right) \delta(1-x)
\Bigg]
+
\frac{\alpha_s}{4 \pi} n_f
\Bigg[
-\frac{2}{3} \left(-\frac{1}{[1-x]}_++x^2+x+1\right)-\frac{10}{9} \delta(1-x)
\Bigg] + \cO(\alpha_s^2) \,,
\nn\\
\frac{1}{2} H_q^{\rm h} = &\ \frac{\alpha_s}{4 \pi} n_f
\Bigg[
-7 x^2+\left(2 x^2-2 x+1\right) (2 \ln (1-x)+4 \ln (x))+4 x
\Bigg] + \cO(\alpha_s^2) \,.
\end{align}
For simplicity we have set $\mu = Q$. Logarithms can be recovered from the RG equation in Eq.~\eqref{eq:hard_evo}.

\end{widetext}


\end{document}